\documentclass[a4paper,fleqn]{cas-dc-arxiv}

\usepackage[numbers]{natbib}
\usepackage{lineno,hyperref}

\usepackage{epsfig}

\usepackage{amssymb}
\usepackage{amsthm}

\usepackage{subfigure}
\usepackage{graphics}
\usepackage{multicol}
\usepackage{graphicx}

\modulolinenumbers[5]

\def\tsc#1{\csdef{#1}{\textsc{\lowercase{#1}}\xspace}}
\tsc{WGM}
\tsc{QE}
\tsc{EP}
\tsc{PMS}
\tsc{BEC}
\tsc{DE}

\begin{document}
\let\WriteBookmarks\relax
\def\floatpagepagefraction{1}
\def\textpagefraction{.001}
\shorttitle{Influence of crosslinkers and magnetic particle distribution on the magnetic properties of MFs}
\shortauthors{D. Mostarac et~al.}

\title [mode = title]{The influence of crosslinkers and magnetic particle distribution along the filament backbone on the magnetic properties of supracolloidal linear polymer-like chains}



\author[1]{Deniz Mostarac}
\cormark[1]
\address[1]{Computational Physics, University of Vienna, Sensengasse 8, Vienna, Austria.}
\ead{deniz.mostarac@univie.ac.at}
\credit{TBD}

\author[1]{Leo Vaughan}
\credit{TBD}

\author[2,3]{Pedro A. S\'anchez}[orcid=0000-0003-0841-6820]
\ead{r.p.sanchez@urfu.ru}
\address[2]{Ural Federal University, 51 Lenin av., Ekaterinburg, 620000, Russian Federation.}
\address[3]{Institute of Ion Beam Physics and Materials Research, Helmholtz-Zentrum Dresden-Rossendorf e.V., D-01314 Dresden, Germany.}
\credit{TBD}

\author[1,2]{Sofia S. Kantorovich}[orcid=0000-0001-5700-7009]
\ead{sofia.kantorovich@univie.ac.at}

\credit{TBD} 

\cortext[cor1]{Corresponding author}

\begin{abstract}
Diverse polymer crosslinking techniques allow the synthesis of linear polymer-like structures whose monomers are colloidal particles. In the case where all or part of these colloidal particles are magnetic, one can control the behaviour of these supracolloidal polymers, known as magnetic filaments (MFs), by applied magnetic fields. However, the response of MFs strongly depends on the crosslinking procedure. In the present study, we employ Langevin dynamics simulations to investigate the influence of the type of crosslinking and the distribution of magnetic particles within MFs on their response to an external magnetic field. We found that if the rotation of the dipole moment of particles is not coupled to the backbone of the filament, the impact of the magnetic content is strongly decreased.
\end{abstract}


\maketitle

\sloppy

\section{Introduction}
The field of soft matter and the idea of smart, soft matter materials has advanced considerably since magnetic fluids were first synthesised \cite{resler1964magnetocaloric}. Soft materials responsive to magnetic fields can be made by combining magnetic micro- or nanoparticles (MNPs) with polymers. Over the years, this idea has grown into a large number of synthetic soft matter systems \cite{2009-odenbach,zrinyi1998kinetics}. Among these systems, magnetic filaments (MFs) \cite{Dreyfus_2005,2008-benkoski}, first synthesised as micron-sized magnetic-filled paramagnetic latex beads forming chains \cite{1998-furst,1999-furst}, remain without recipes for finely tuning their magnetic response, and are the subject of this manuscript.

MFs can nowadays be found in diverse applications \cite{WANG_2011,wang2014multifunctional,cebers2016flexible,cai2018fluidic}. They can serve as artificial swimmers \cite{2005-dreyfus,2008-erglis-mh}, for cellular engineering \cite{fayol2013use,gerbal2015refined} and for bio-mimetic cilia designs \cite{evans2007magnetically,erglis2011three}. Theoretically, MFs have mostly been explored in bulk \cite{2003-cebers,kuznetsov2019equilibrium}. In all these studies there is an agreement concerning the fact that the behaviour of MFs strongly depends on their rigidity and overall magnetic properties. In turn, it is reasonable to conjecture that such properties should strongly depend on the type of crosslinks bonding the particles and the distribution of magnetic properties of the latter.

To the best of our knowledge, there are currently no comparative studies of nanoscale, ferromagnetic MFs, diverse in terms of distribution of magnetic particles along the filament and crosslinking rigidity, while exposed to an external magnetic field. In this paper, by performing Molecular Dynamics (MD) simulations with the {ESPResSo} software package \cite{2013-arnold}, we attempt to bridge this gap. We employ two crosslinking models for MFs, representing distinct crosslinking scenarios as described in Section \ref{sec:model}. In Section \ref{sec:res-disc} we present a comparative analysis of the equilibrium magnetic properties of MFs in constant, homogeneous magnetic fields for both crosslinking models and three different distributions of magnetic particles. In the last section we provide a brief summary of the results and an outlook.

\section{Model and simulation details}\label{sec:model}

\subsection{Non-bonding interactions}
We consider MFs to be made of $L=20$ or $L=60$ monodisperse colloidal particles, modelled as identical spherical beads with characteristic dimensionless diameter $\sigma = 1$ and mass $m=1$. Such particles can be either nonmagnetic or ferromagnetic. The latter carry a point magnetic dipole moment $\vec{\mu}$ located at their centres. We account for the long-range magnetic interparticle interactions by means of the conventional dipole-dipole pair potential:
\begin{equation}
U_{dd}(\vec r_{ij})=\frac{\vec{\mu}_{i}\cdot\vec{\mu}_{j}}{r^{3}}-\frac{3\left[\vec{\mu}_{i}\cdot\vec{r}_{ij}\right]\left[\vec{\mu}_{j}\cdot\vec{r}_{ij}\right]}{r^{5}},
\label{eq:dipdip}
\end{equation}
where $\vec \mu_i$ and $\vec \mu_j$ are the respective dipole moments of the interacting particles, $\vec r_{ij} = \vec r_i - \vec r_j$ is the displacement vector connecting their centres and $r=\left \| \vec r _{ij}\right \|$. 

Furthermore, we consider Zeeman interactions between the external magnetic field $\vec{H}$ and any magnetic particle with dipole moment $\vec \mu_i$:
\begin{equation}
U_H(\vec{H},\vec{\mu_i})=-\vec{H}\cdot \vec{\mu_i}.
\label{eq:zee}
\end{equation}

The soft core interaction between any two colloids is given by the Weeks-Chandler-Andersen pair potential (WCA) \cite{weeks1971role}, that in dimensionless units we define as:
\begin{equation}
U_{W C A}(r)=\left\{\begin{array}{ll}{4\left[ r^{-12}-r^{-6} +\frac{1}{4}\right],} & {r \leqslant 2^{1/6}} \\ {0,} & {r > 2^{1/6}}\end{array}\right.
\label{eq:wca}
\end{equation}
\subsection{Crosslinking and sequencing}
\begin{figure}
	\centering
	\includegraphics[width=0.3 \textwidth]{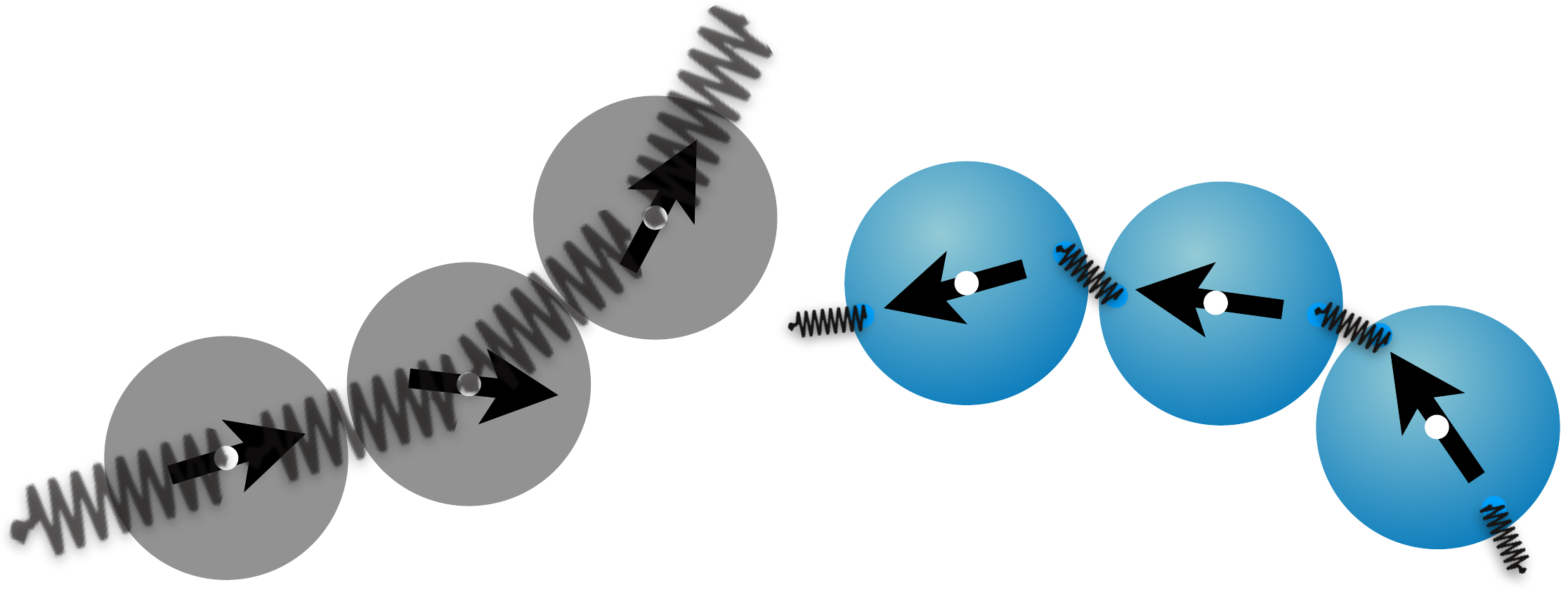}
	\caption{Scheme of the two crosslinking approaches considered in this study: \textit{plain} on the left, \textit{constrained} on the right. Particle magnetic moments are denoted with black arrows. On the left, magnetic particles can rotate without bonding energy penalty, as the crosslinking springs are attached to their centres. On the right, crosslinking springs are attached to the surface of the particles, thus relative particle rotations lead to spring deformation.}
	\label{fig:cross}
\end{figure}

We introduce two types of crosslinking, represented by elastic springs with different attachment points to the crosslinked particles, as shown in Fig. \ref{fig:cross}. In the first type, here referred to as the \textit{plain} crosslinking model (gray particles, left figure), springs are attached to the centre of the particles, so that their rotations are not constrained by the crosslinking. In the second, or \textit{constrained} crosslinking model (green particles, right figure), springs are attached to the surface of the particles (that is, at a distance 0.5 from their centers), coupling the filament backbone to the orientation of the dipoles.
\begin{figure}
	\centering
	\includegraphics[width=0.7 \columnwidth]{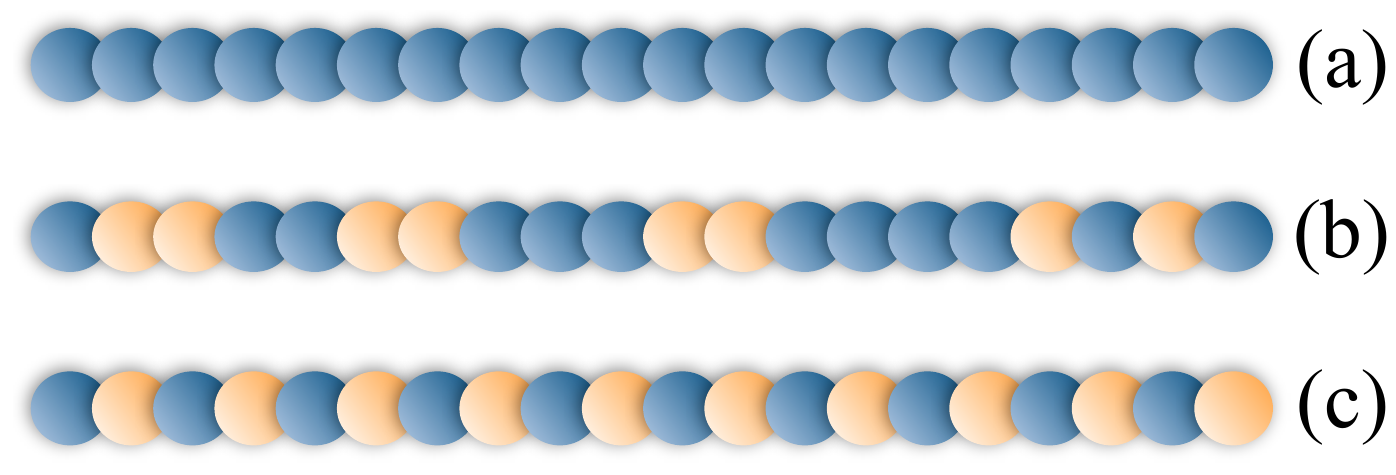}
	\caption{Three types of particle sequences analysed in this study. Magnetic particles are shown in dark blue, nonmagnetic in light yellow: (a) \textit{all} particles are magnetic; (b) arbitrarily selected aperiodic \textit{sequence} of magnetic/nonmagnetic particles; (c) \textit{alternating} configuration.}
	\label{fig:sequence}
\end{figure}
Both crosslinking mechanisms have already been used to study properties of MFs in which all particles were magnetic \cite{2015-sanchez-sm1}. Here, instead, we use such a filament as a reference to compare with two other inhomogeneous distributions of magnetic particles, shown in Fig. \ref{fig:sequence}. Throughout the paper, we will address these configurations as \textit{all}, when referring to conventional filaments in which all beads are magnetic (a); as \textit{sequence} when we talk about the arbitrarily selected aperiodic configuration (b); and as \textit{alternating} when we denote the configuration with alternating magnetic and nonmagnetic beads (c). For the configuration of NPs we denote as "sequence", the number of magnetic NPs is $N_{mag}=12$ for $L=20$ and $N_{mag}=36$ for $L=60$. For the configuration of NPs we refer to as "alternating", the number of magnetic NPs is $N_{mag}=10$ for $L=20$ and $N_{mag}=30$ for $L=60$.

\subsection{Simulation protocol}
We performed MD simulations of single filaments using a Langevin thermostat and open boundaries. The dimensionless temperature was fixed to $T=1$. Due to the relatively low amount of magnetic particles in the system, magnetic interactions were calculated by direct sum. Simulations started with an equilibration cycle of $10^{7}$ integration steps, after which the external magnetic field was switched on and a production run of $3.5\cdot 10^5$ integration steps was carried out. Time step was fixed to $\delta t = 10^{-2}$. Measurements were sampled every $3\cdot10^3$ integration steps. Results presented here are based on averages over twenty independent runs with different initial configurations.

\section{Results}\label{sec:res-disc}
\begin{figure*}
	\centering
\subfigure[]{\label{fig:M-l1-pcl-l20}\includegraphics[width=0.245 \textwidth]{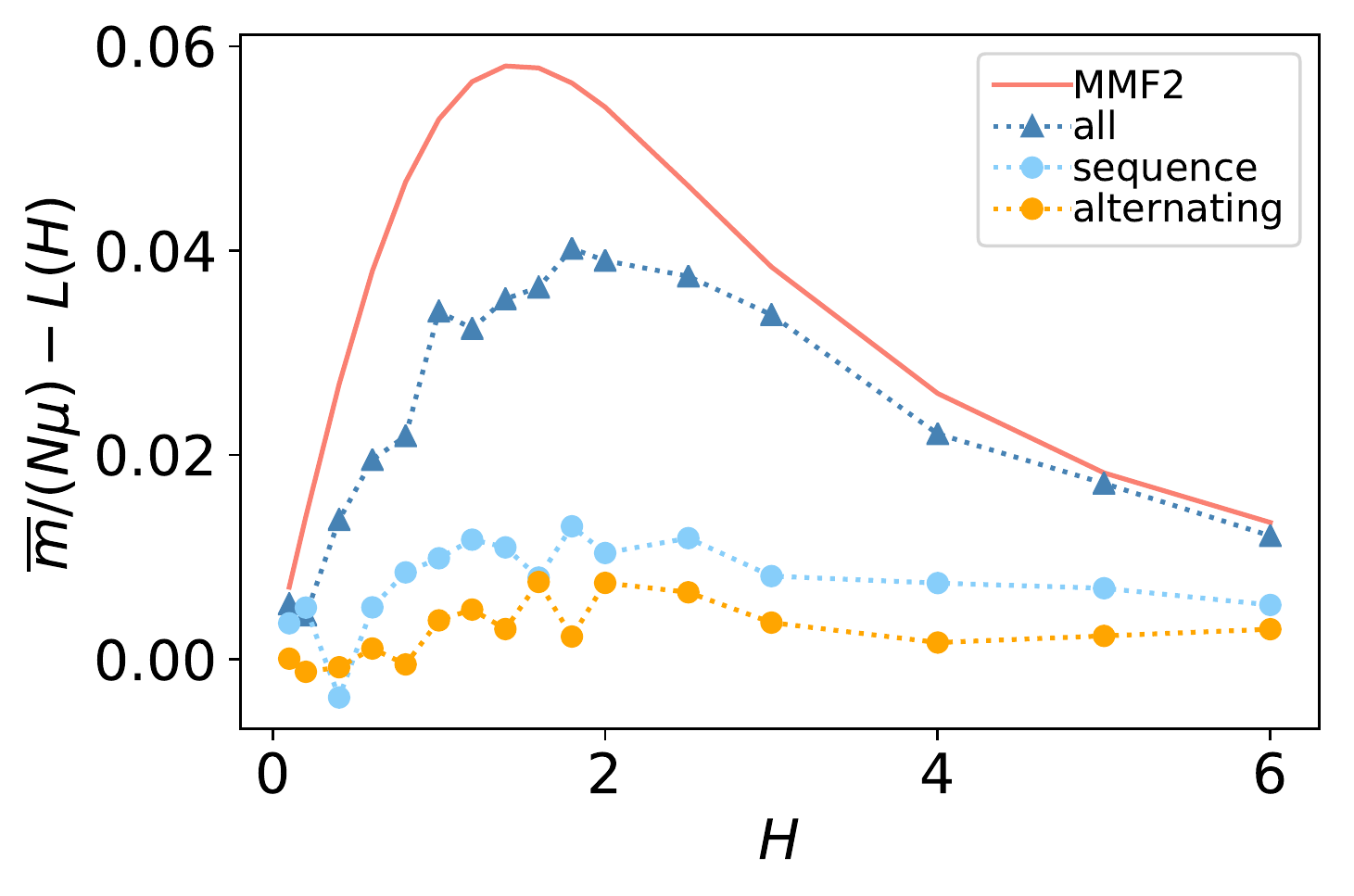}}
\subfigure[]{\label{fig:M-l3-pcl-l20}\includegraphics[width=0.245 \textwidth]{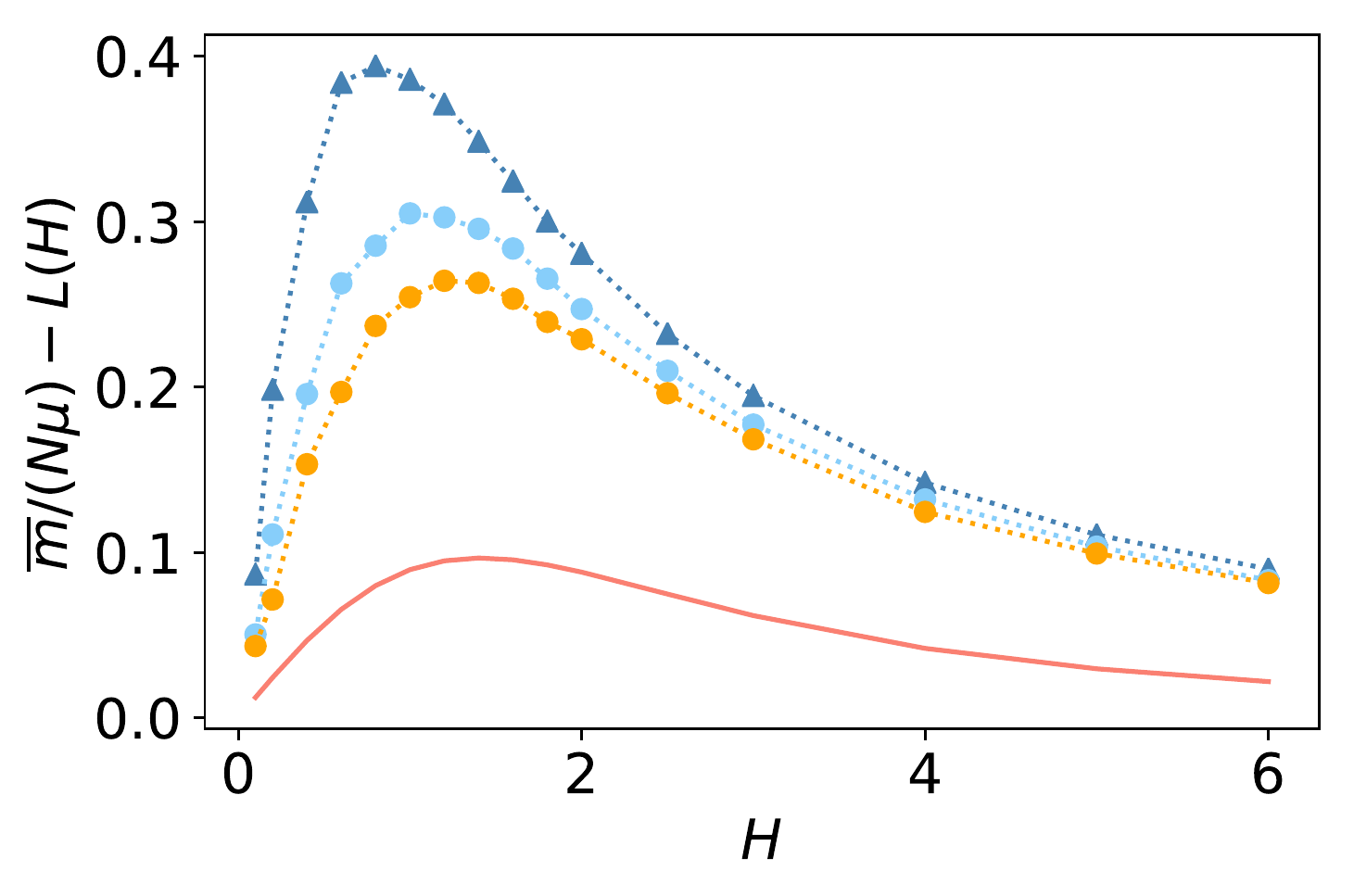}}
\subfigure[]{\label{fig:M-l1-ccl-l20}\includegraphics[width=0.245 \textwidth]{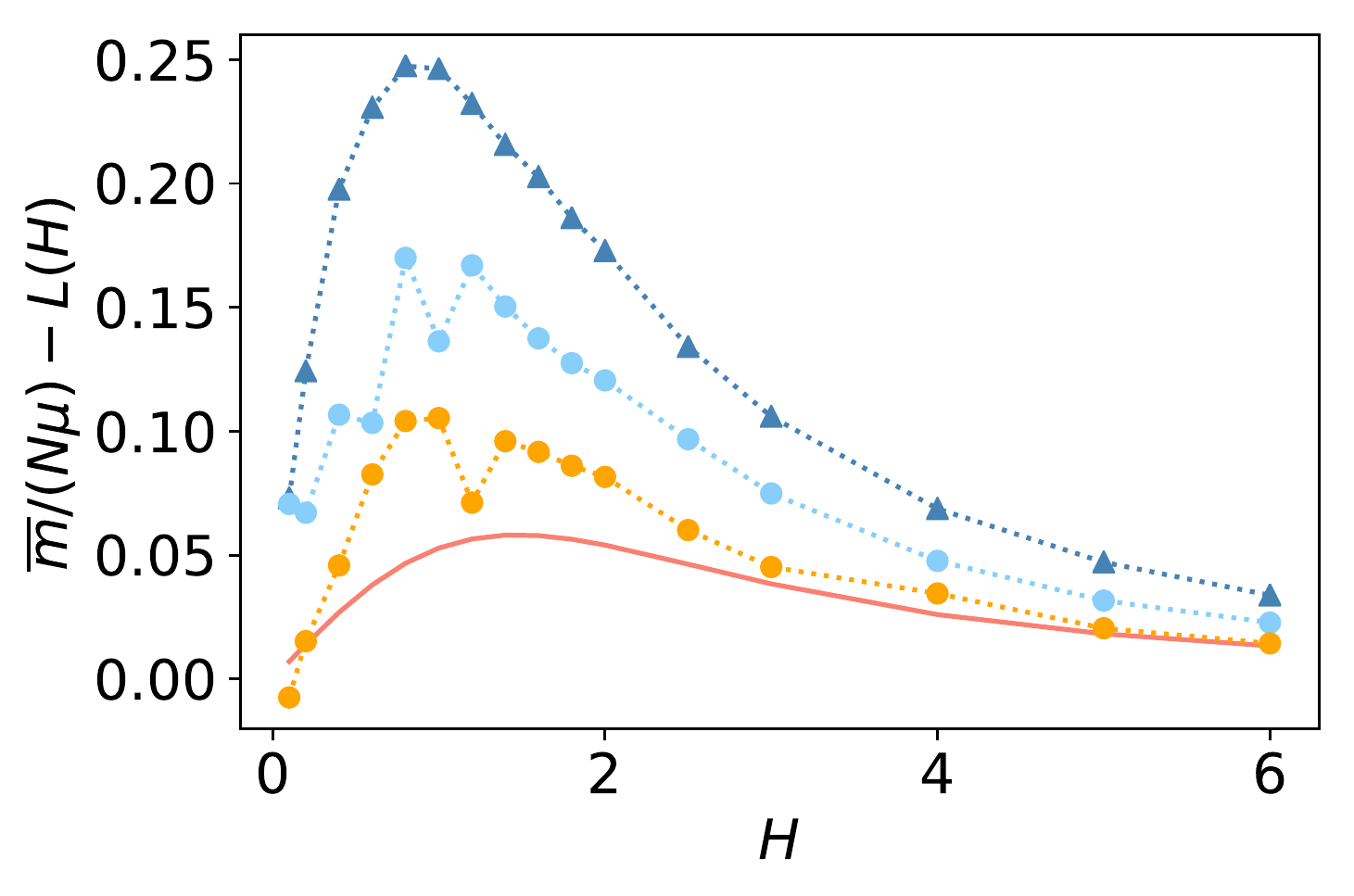}}
\subfigure[]{\label{fig:M-l3-ccl-l20}\includegraphics[width=0.245 \textwidth]{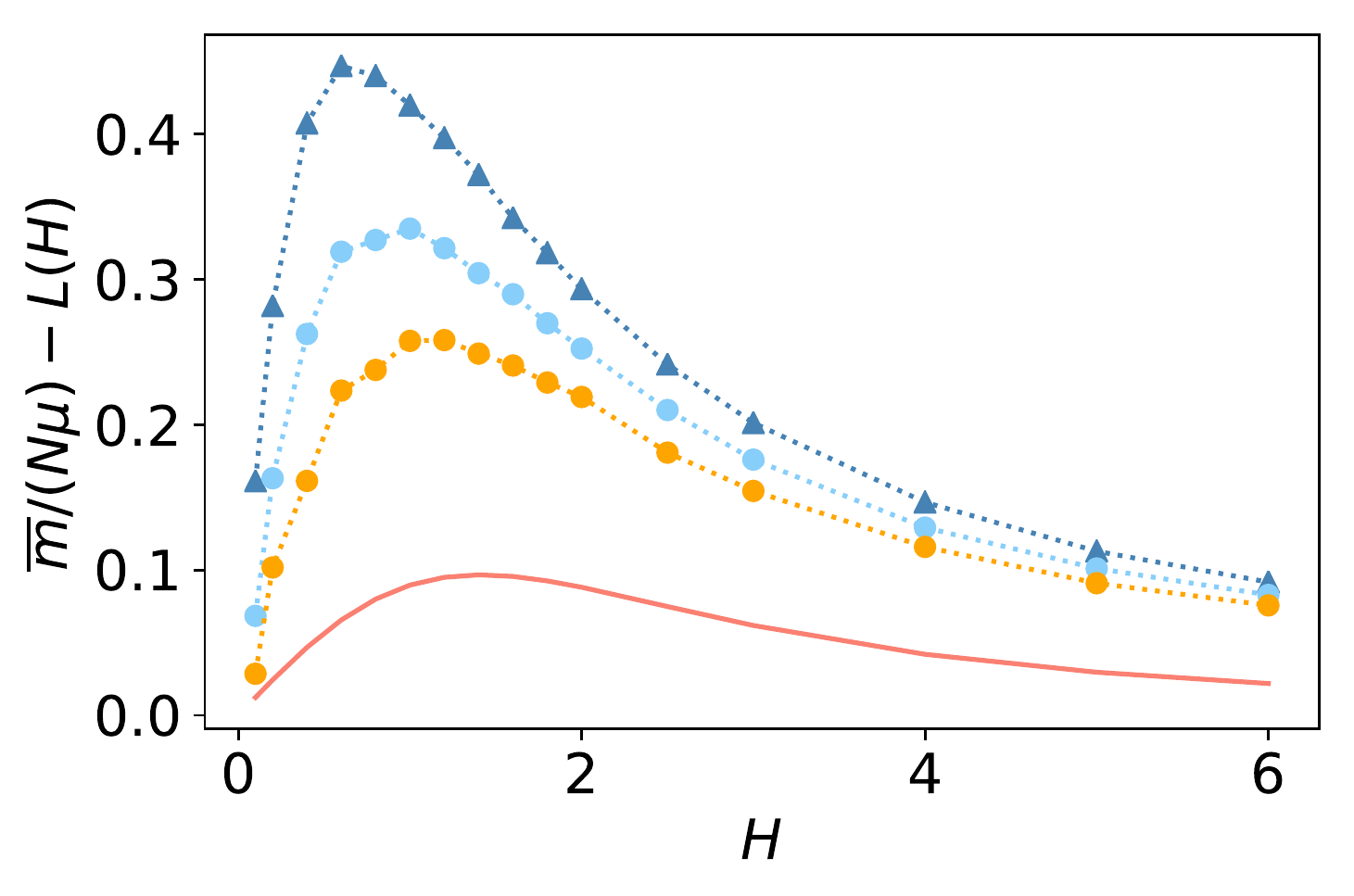}}
 \subfigure[]{\label{fig:M-l1-pcl-l60}\includegraphics[width=0.245 \textwidth]{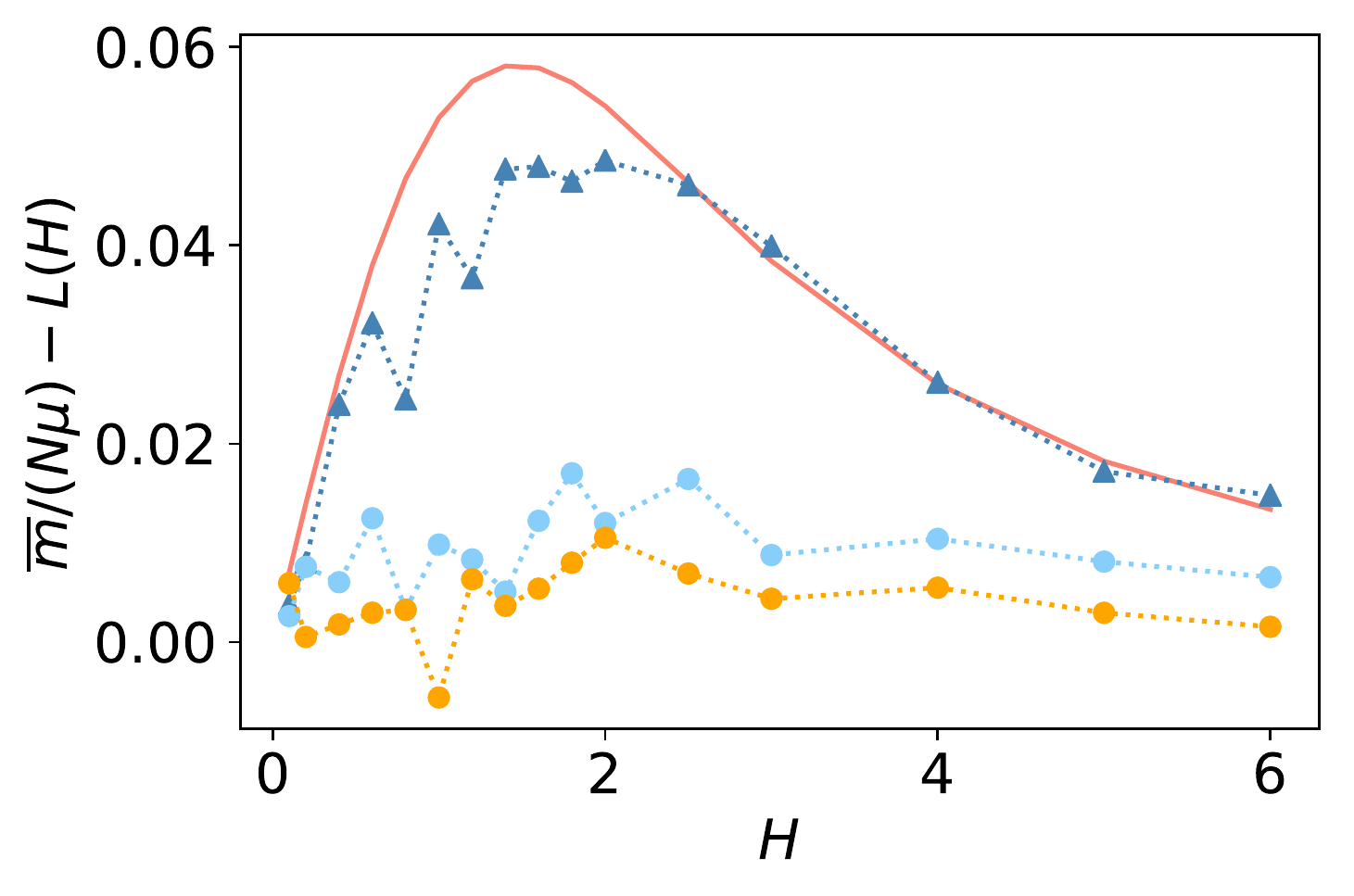}}
\subfigure[]{\label{fig:M-l3-pcl-l60}\includegraphics[width=0.245 \textwidth]{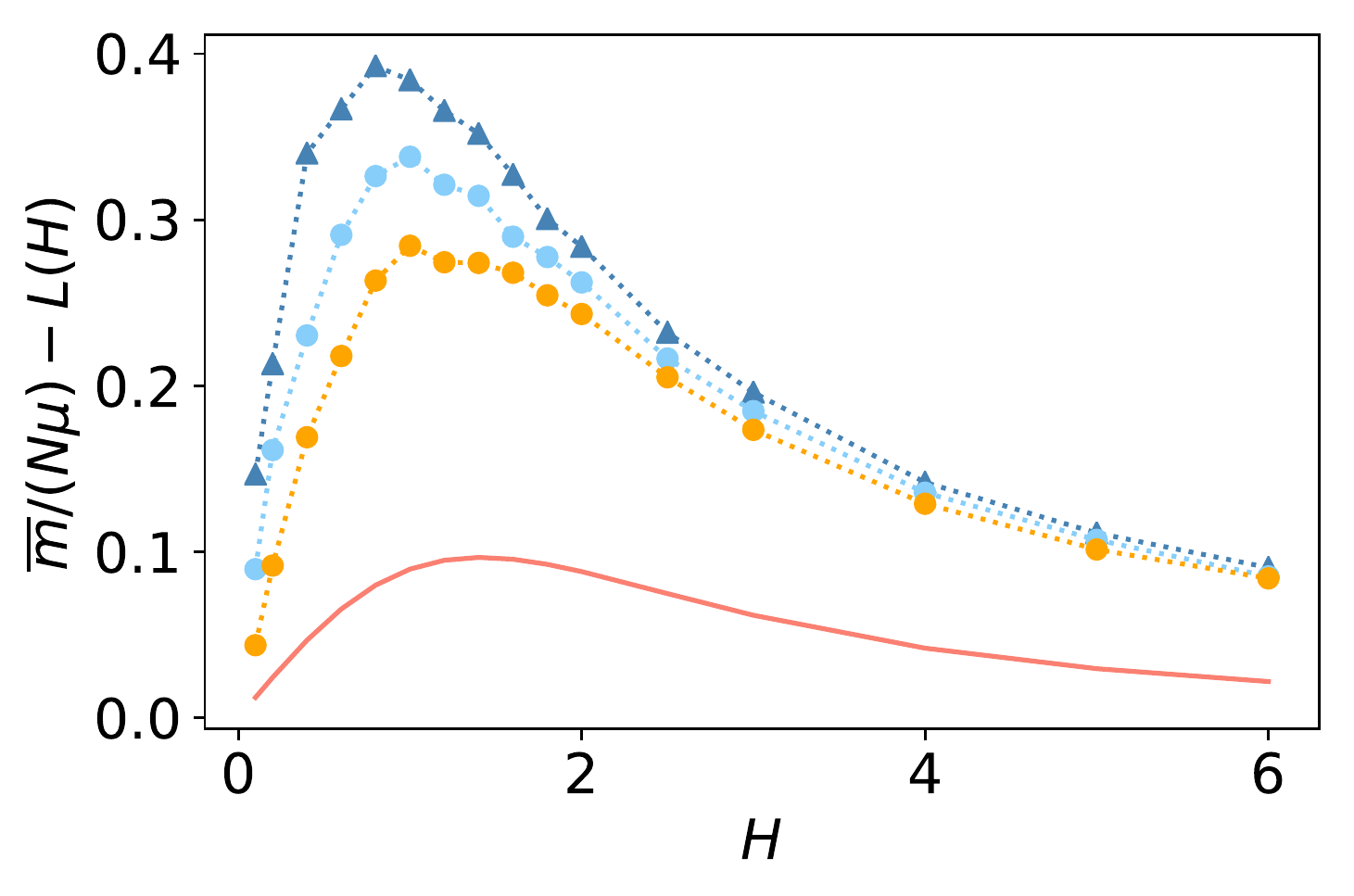}}
\subfigure[]{\label{fig:M-l1-ccl-l60}\includegraphics[width=0.245 \textwidth]{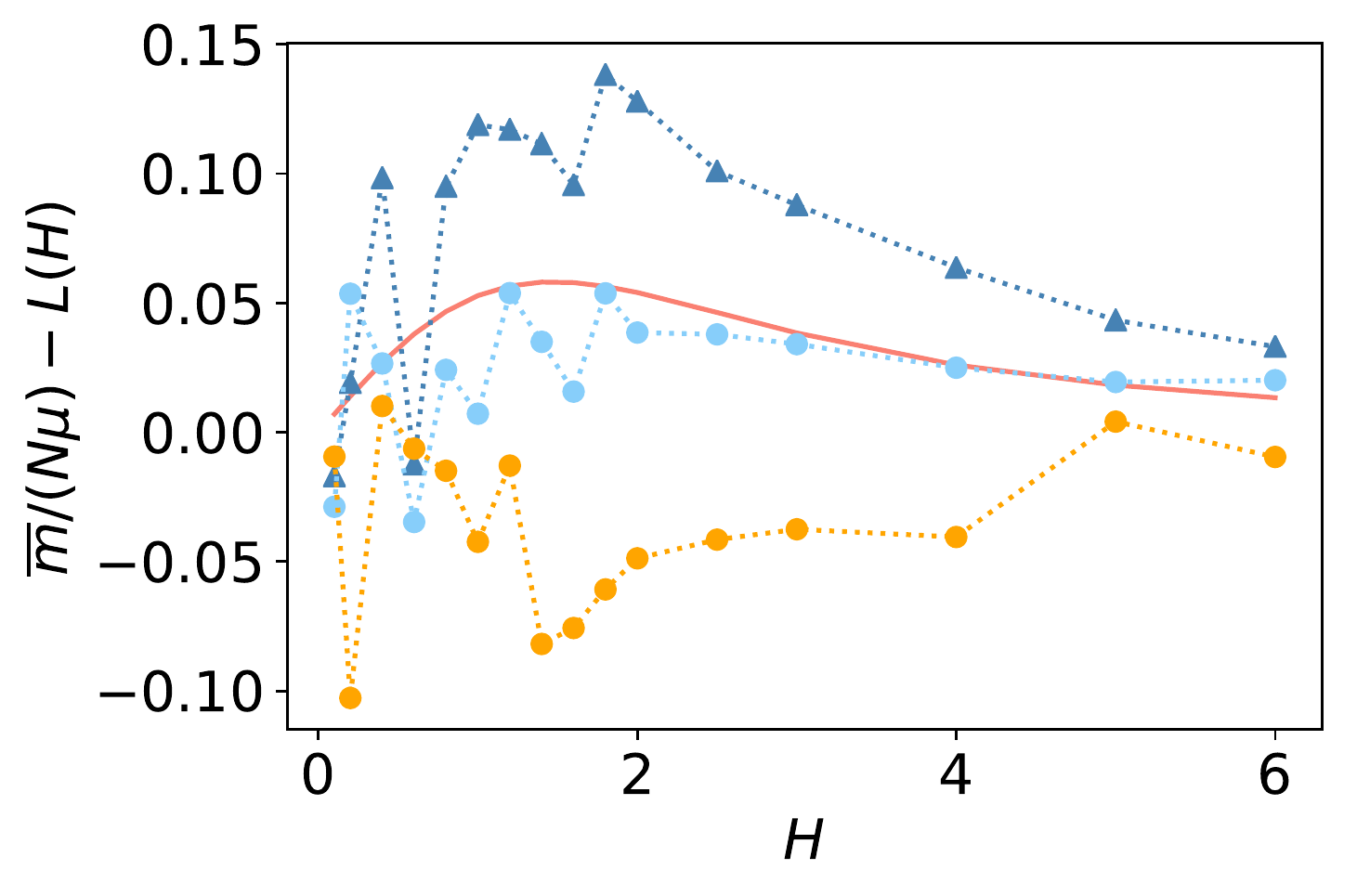}}
\subfigure[]{\label{fig:M-l3-ccl-l60}\includegraphics[width=0.245 \textwidth]{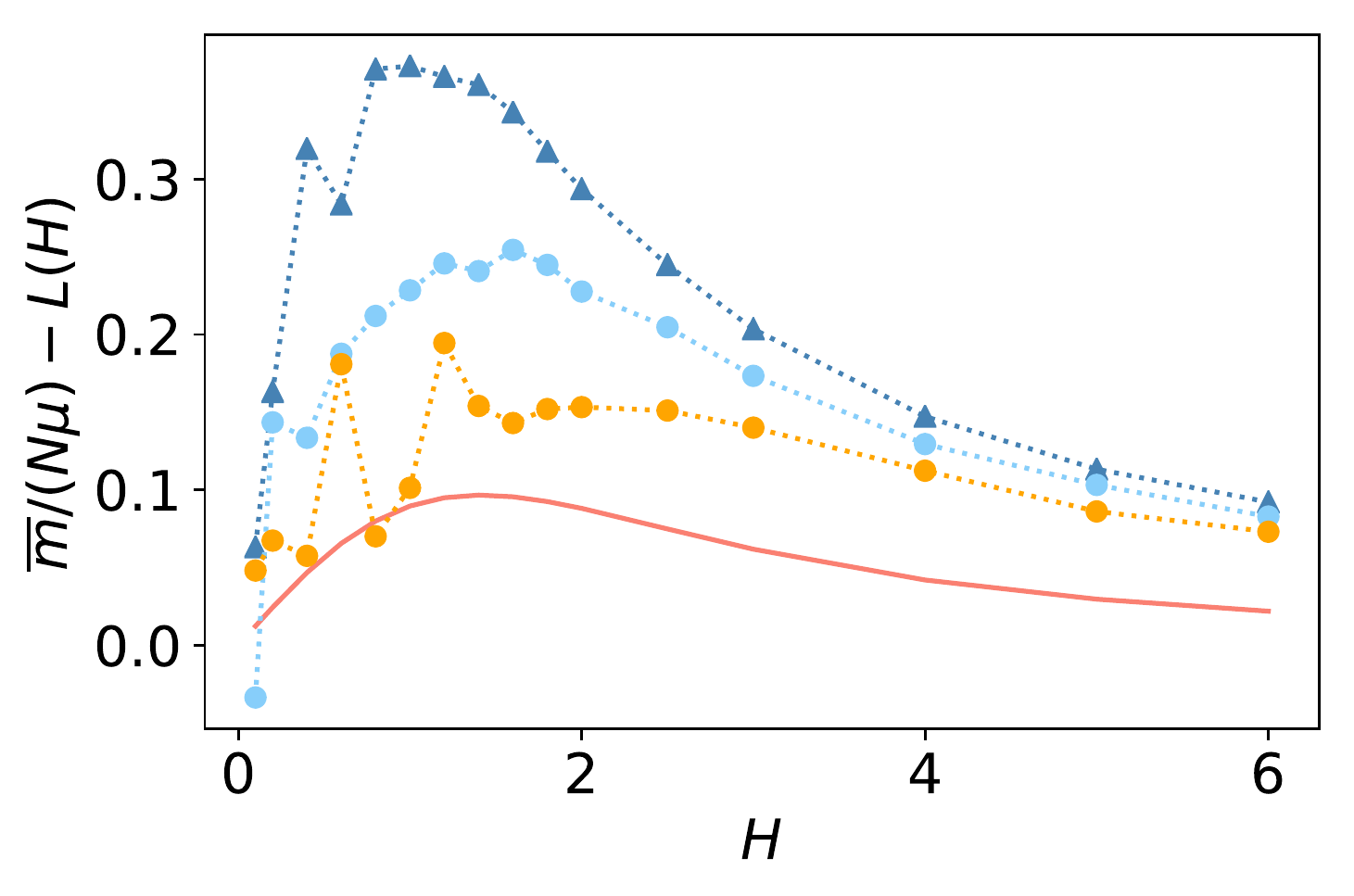}}
	\caption{The difference between the normalised total magnetic moment of the filament, $N_{mag}/(N\mu)$ and the Langevin function, $L(H)$, for different values of $\mu^2$ and type of crosslinking. $N$ is the number of magnetic beads while $\mu$ is the magnetic moment of a bead. In all subfigures, MMF2 is plotted with full, salmon colour lines. The sequencing is explained in the legend. In (a)--(d) $L = 20$; in (e)--(h) $L=60$. In (a) and (e) the magnetisation is for MFs with plain crosslinking and $\mu^2=1$; in (b) and (f) the magnetisation is for MFs with plain crosslinking and $\mu^2=3$; (c) and (g) are for MFs with constrained crosslinking and $\mu^2=1$; (d) and (h) are for MFs with constrained crosslinking and $\mu^2=3$.}
	\label{fig:magn_mom_pp}
\end{figure*}
We start the discussion of the results by presenting, in Fig.~\ref{fig:magn_mom_pp}, the difference between the normalised per particle magnetisation of filaments, $N_{mag}/(N\mu_{max})$ and the Langevin function, $L(H)$, as a function of the dimensionless applied magnetic field, $H$. The Langevin function, represents the magnetisation of a single magnetic nanoparticle, immersed in a homogeneous magnetic field, and is given by:
\begin{equation}\label{eq:langevin}
L(H)=\coth(H)-\dfrac{1}{H}
\end{equation}
The upper row includes the results for $L=20$, whereas the lower row corresponds to $L=60$. Note that filament conformations with $L=60$ were made simply by repeating the motive established for filament topologies with $L=20$, three times, respectively. In this way, we ensure that the ratio of non-magnetic and magnetic particles in a filament conformation remains the same, regardless of filament length. The two leftmost columns are for \textit{plain} crosslinking, the two rightmost ones for \textit{constrained}. Figs.~\ref{fig:M-l1-pcl-l20}, \ref{fig:M-l1-pcl-l60}, \ref{fig:M-l1-ccl-l20} and \ref{fig:M-l1-ccl-l60} correspond to weak magnetic interactions, $\mu^2=\| \vec \mu \|^2=1$, whereas Figs.~\ref{fig:M-l3-pcl-l20}, \ref{fig:M-l3-pcl-l60}, \ref{fig:M-l3-ccl-l20} and \ref{fig:M-l3-ccl-l60} are for moderate magnetic interactions, $\mu^2=3$. Along with the simulation data for different sequences and crosslinking mechanisms, each figure shows the difference between Langevin magnetisation given by eq. \eqref{eq:langevin} and the magnetisation based on the modified mean-field theoretical prediction of the second order (MMF2, shown in salmon),\cite{ivanov2001magnetic} that takes into account the contribution of the dipole interactions between particles into the magnetisation. In order to calculate the effective field in the framework of the MMF2, we use the following approach. The total volume of the system is estimated as the sphere with its radius being equal to the radius of gyration of a filament. In this volume the local volume fraction of magnetic particles is calculated. In Fig.~\ref{fig:magn_mom_pp}, we always use the same volume fraction of magnetic particles obtained for the smallest radius of gyration observed in simulations, as MMF2 is used to underline the impact of crosslinking. Here, comparing the upper and lower rows of Fig.~\ref{fig:magn_mom_pp}, one can say that for both types of crosslinking and for different values of $\mu^2$ there is no qualitative change in the magnetisation if the filament length increases by a factor of three. If we consider the same type of crosslinking and focus on the influence of the value of $\mu^2$, one can notice that for higher values of magnetic moment the difference between the three sequences is amplified (comparing, for example, Figs.~ \ref{fig:M-l1-pcl-l20} and \ref{fig:M-l3-pcl-l20} respectively to Figs.~\ref{fig:M-l1-ccl-l60} and \ref{fig:M-l3-ccl-l60}). The highest magnetisation, as expected, is always observed for fully magnetic MFs, whereas the lowest magnetisation per magnetic particle is found for the alternating sequence. Especially pronounced is the difference between the sequences in the low-field region and for constrained crosslinking: the greater the number of magnetic particles found to be next to each other in the filament, the more correlated they are and thus the higher the initial magnetic susceptibility. Interestingly, independently from the value of $\mu^2$, the difference between the magnetisation curves of the different sequences is always higher for MFs with \textit{constrained} crosslinking. Additionally, the interparticle correlations for $\mu^2 = 1$ are also affected by the crosslinking type: comparing the magnetisation for MFs with \textit{plain} crosslinking in Fig.~\ref{fig:M-l1-pcl-l20} to their counterparts with \textit{constrained} crosslinking in Fig.~\ref{fig:M-l1-ccl-l20}, it is clearly seen that in the first case the magnetisation barely deviates from the analytical predictions of MMF2, whereas for the second case the magnetisation is higher than the MMF2 curve for any sequence. This effect, however, vanishes if $\mu^2$ is sufficiently large for magnetic particles to be correlated due to dipole-dipole forces, and the comparison between magnetisation curves in Figs. \ref{fig:M-l3-pcl-l20} and \ref{fig:M-l3-ccl-l20}, as well as in Figs.~\ref{fig:M-l3-pcl-l60} and \ref{fig:M-l3-ccl-l60}, reveals no significant differences. Note that the noise in some curves is related to the statistical error.

\begin{figure*}
	\centering
\subfigure[]{\label{fig:hist-l1-pcl-l20-h02}\includegraphics[width=0.245 \textwidth]{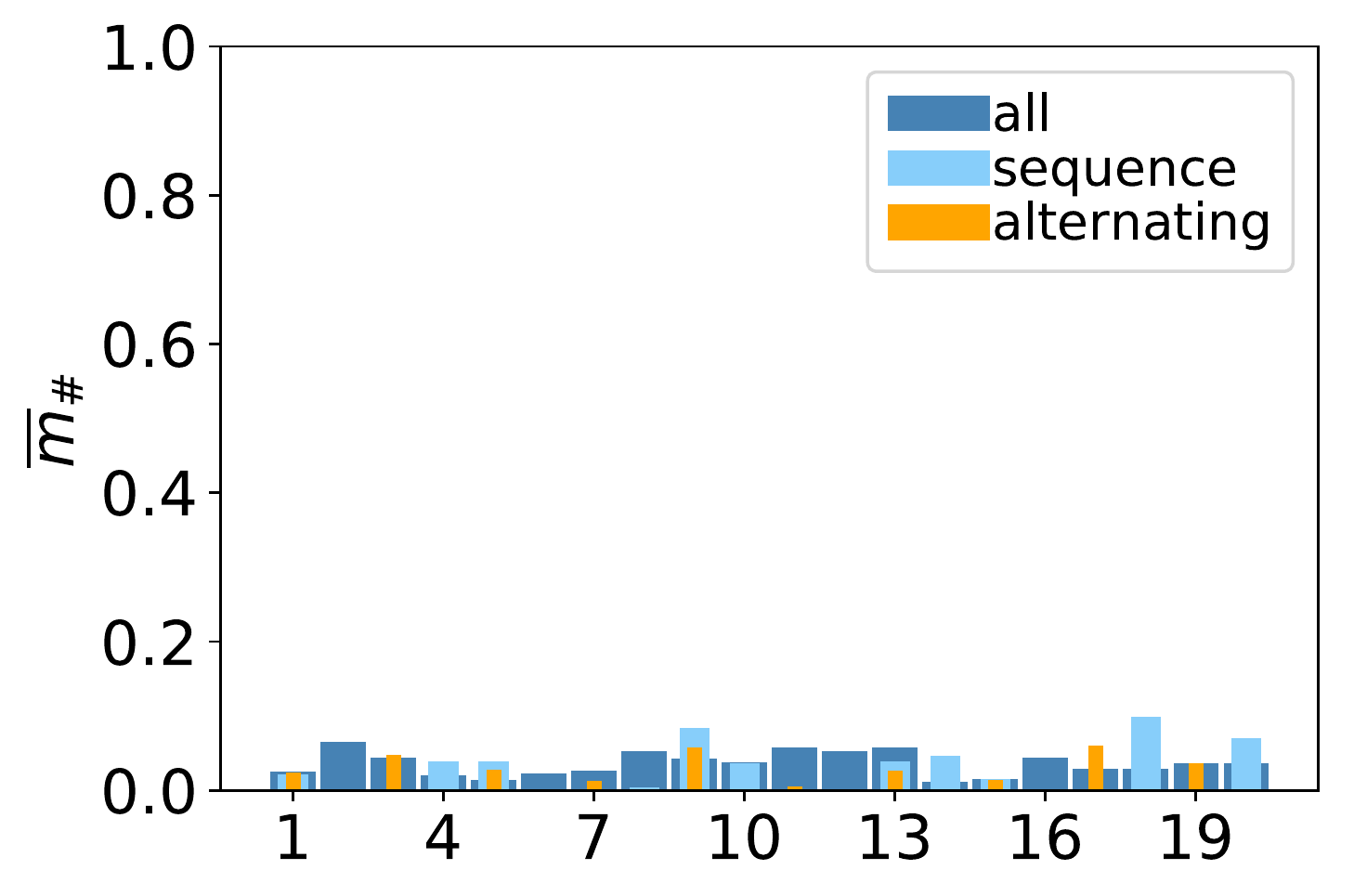}}
\subfigure[]{\label{fig:hist-l1-pcl-l20-h60}\includegraphics[width=0.245 \textwidth]{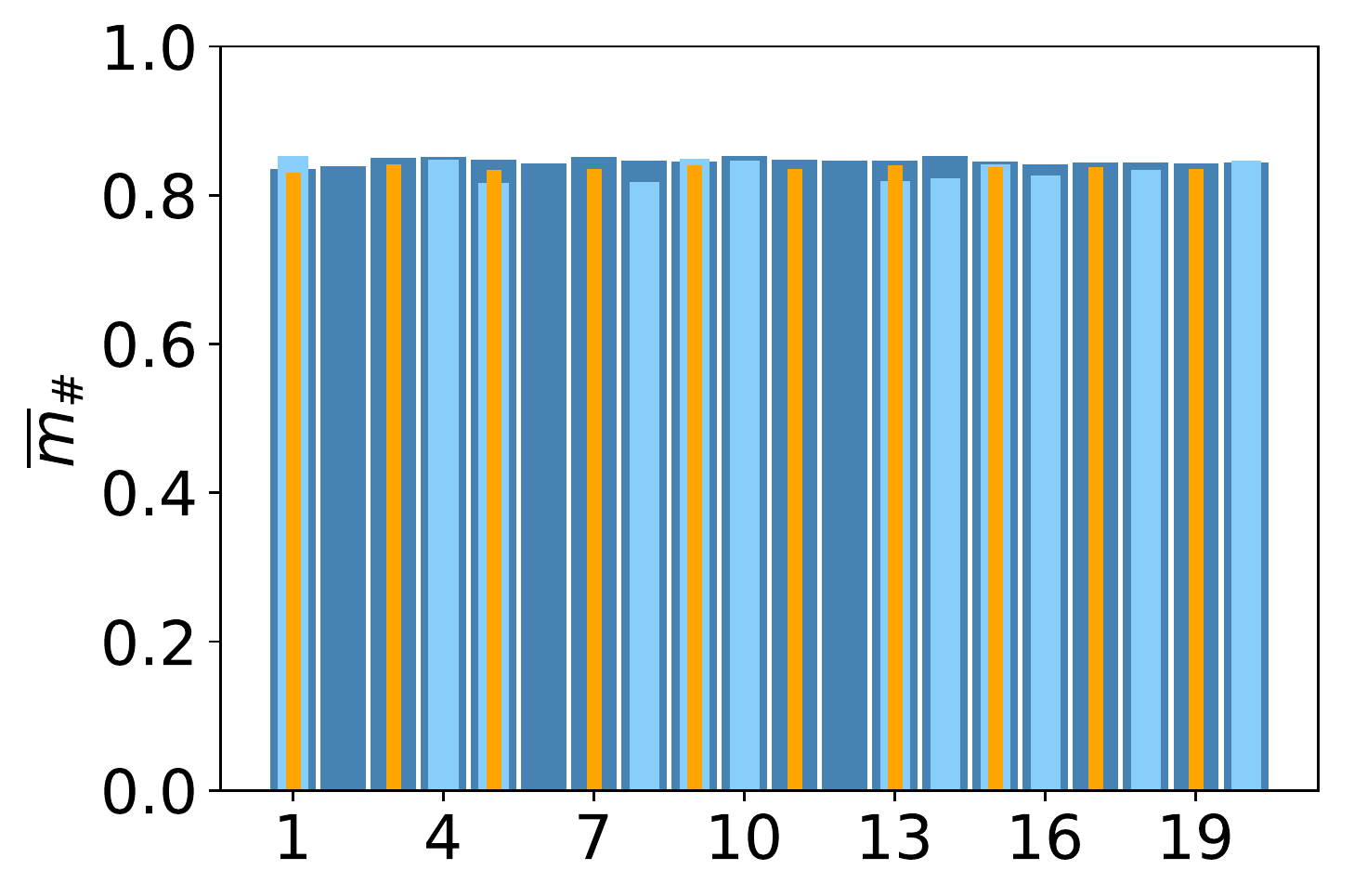}}
\subfigure[]{\label{fig:hist-l1-pcl-l60-h02}\includegraphics[width=0.245 \textwidth]{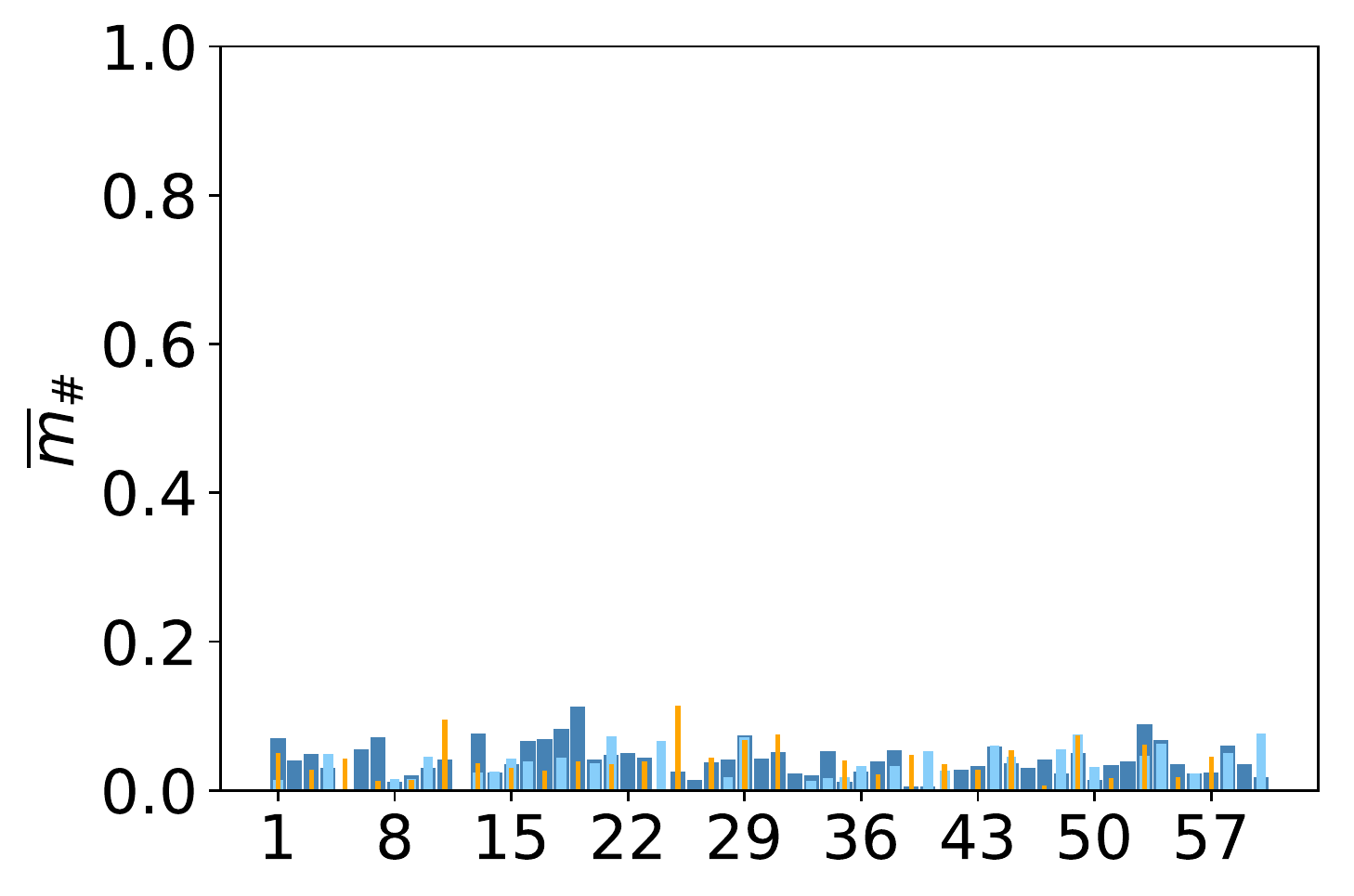}}
\subfigure[]{\label{fig:hist-l1-pcl-l60-h60}\includegraphics[width=0.245 \textwidth]{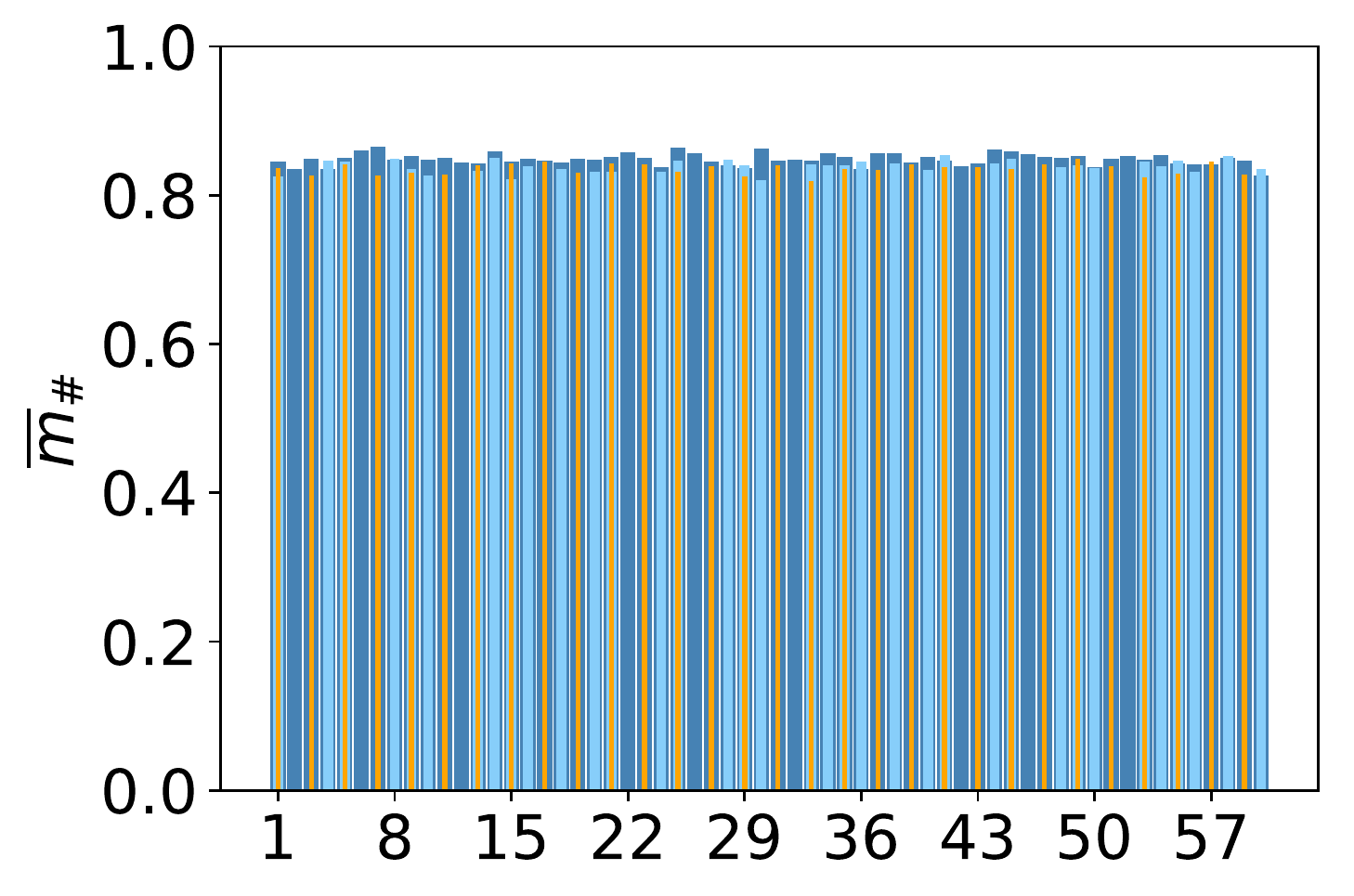}}
\subfigure[]{\label{fig:hist-l3-pcl-l20-h02}\includegraphics[width=0.245 \textwidth]{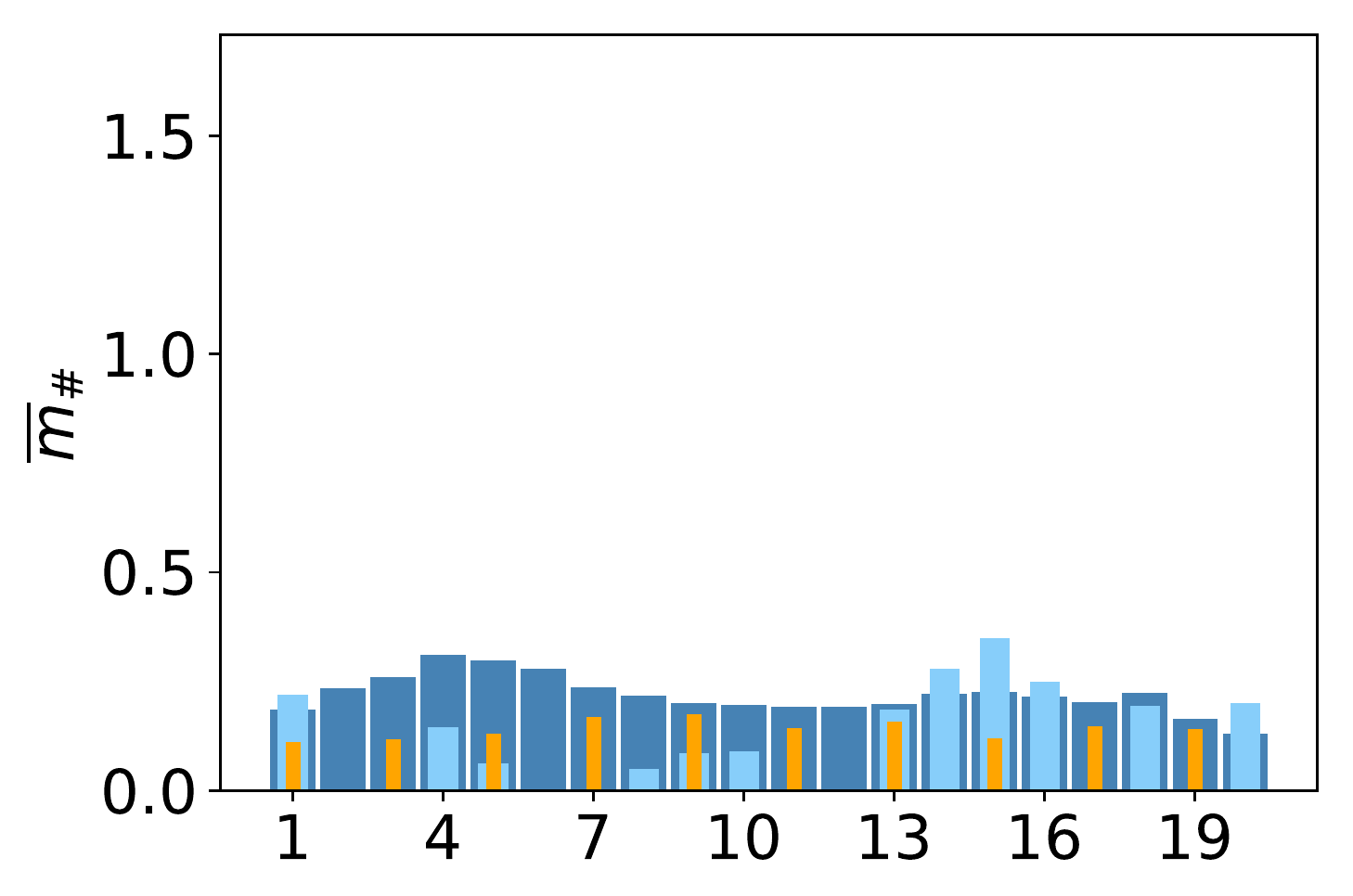}}
\subfigure[]{\label{fig:hist-l3-pcl-l20-h60}\includegraphics[width=0.245 \textwidth]{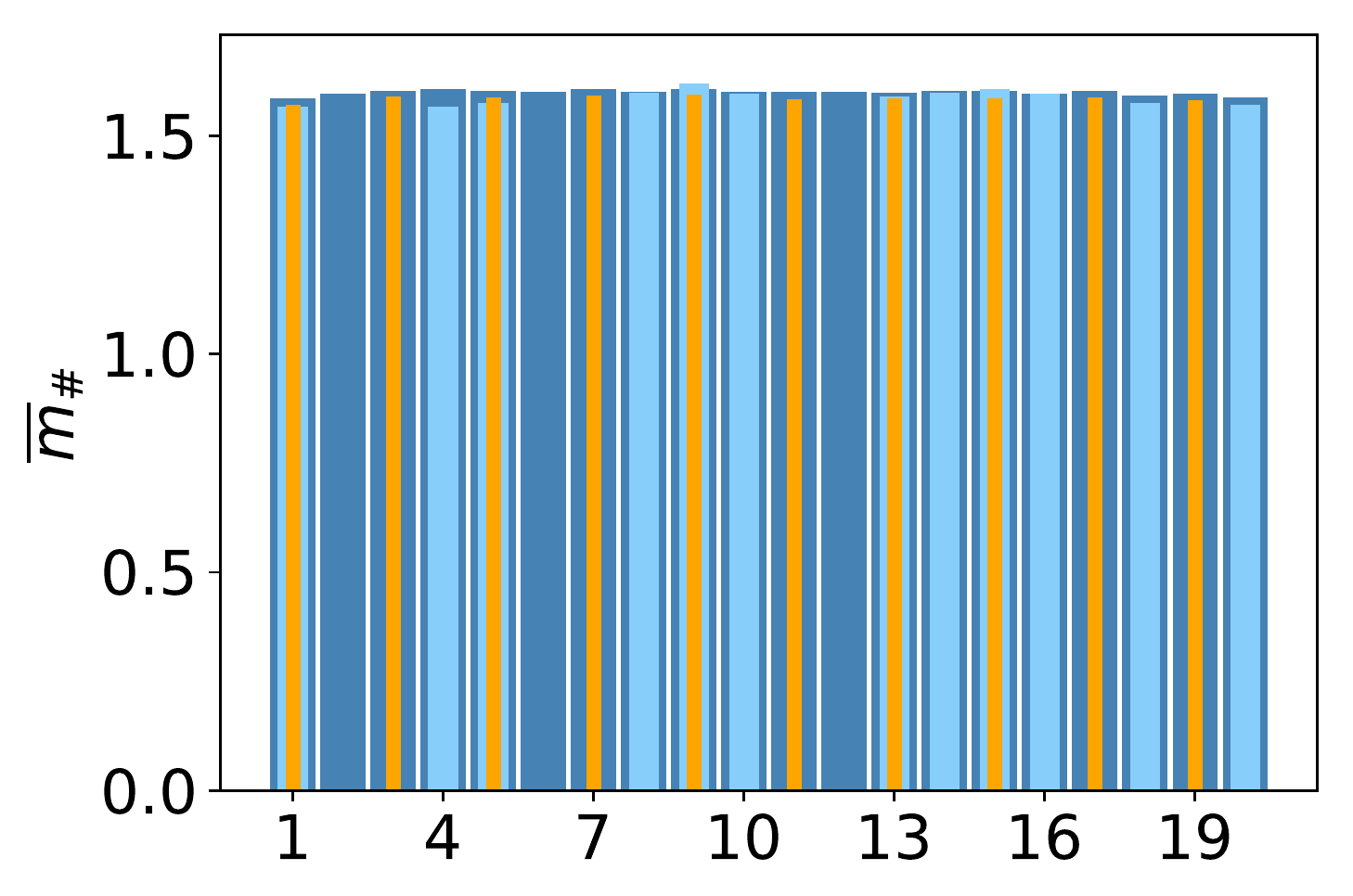}}
\subfigure[]{\label{fig:hist-l3-pcl-l60-h02}\includegraphics[width=0.245 \textwidth]{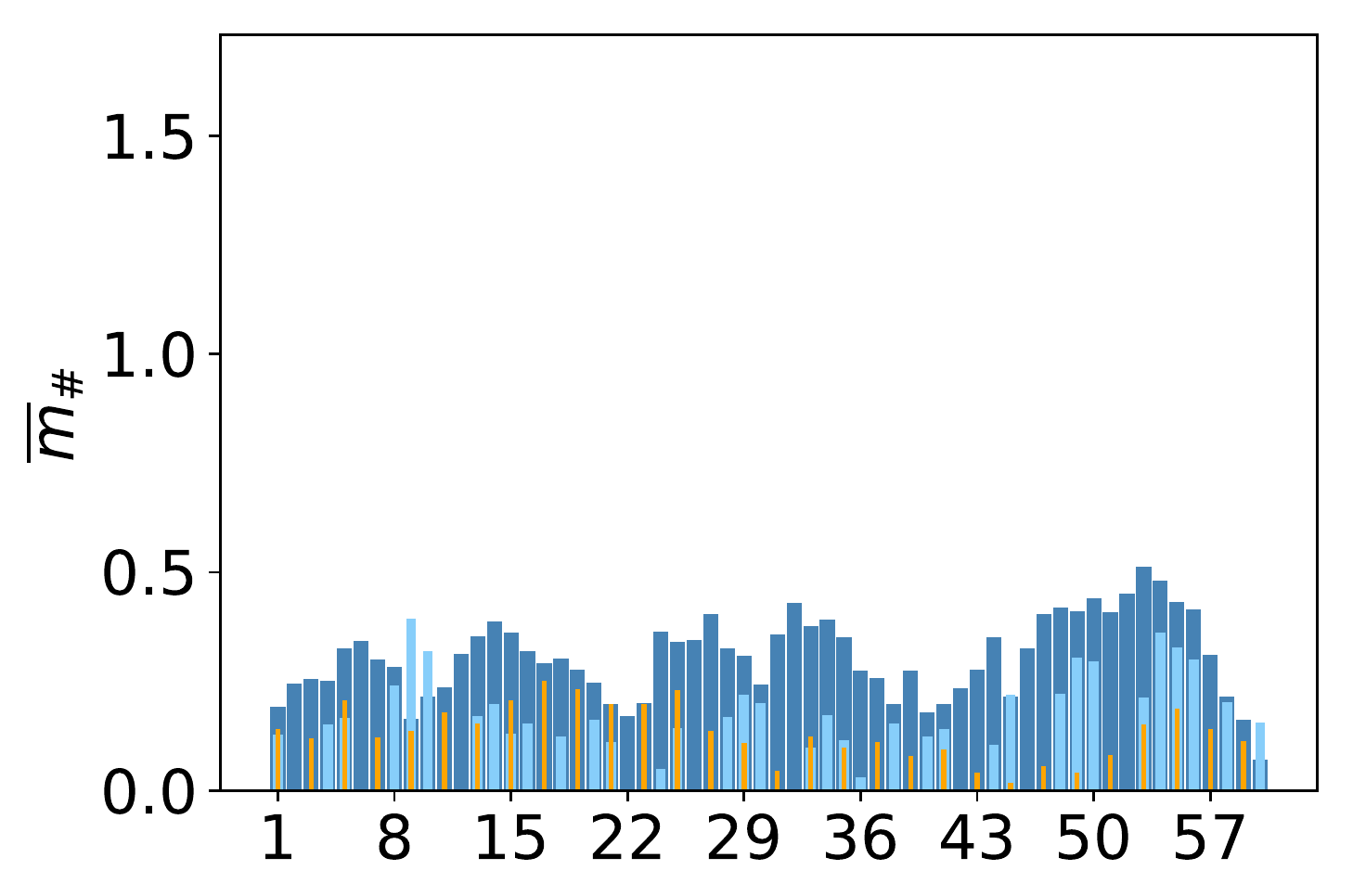}}
\subfigure[]{\label{fig:hist-l3-pcl-l60-h60}\includegraphics[width=0.245 \textwidth]{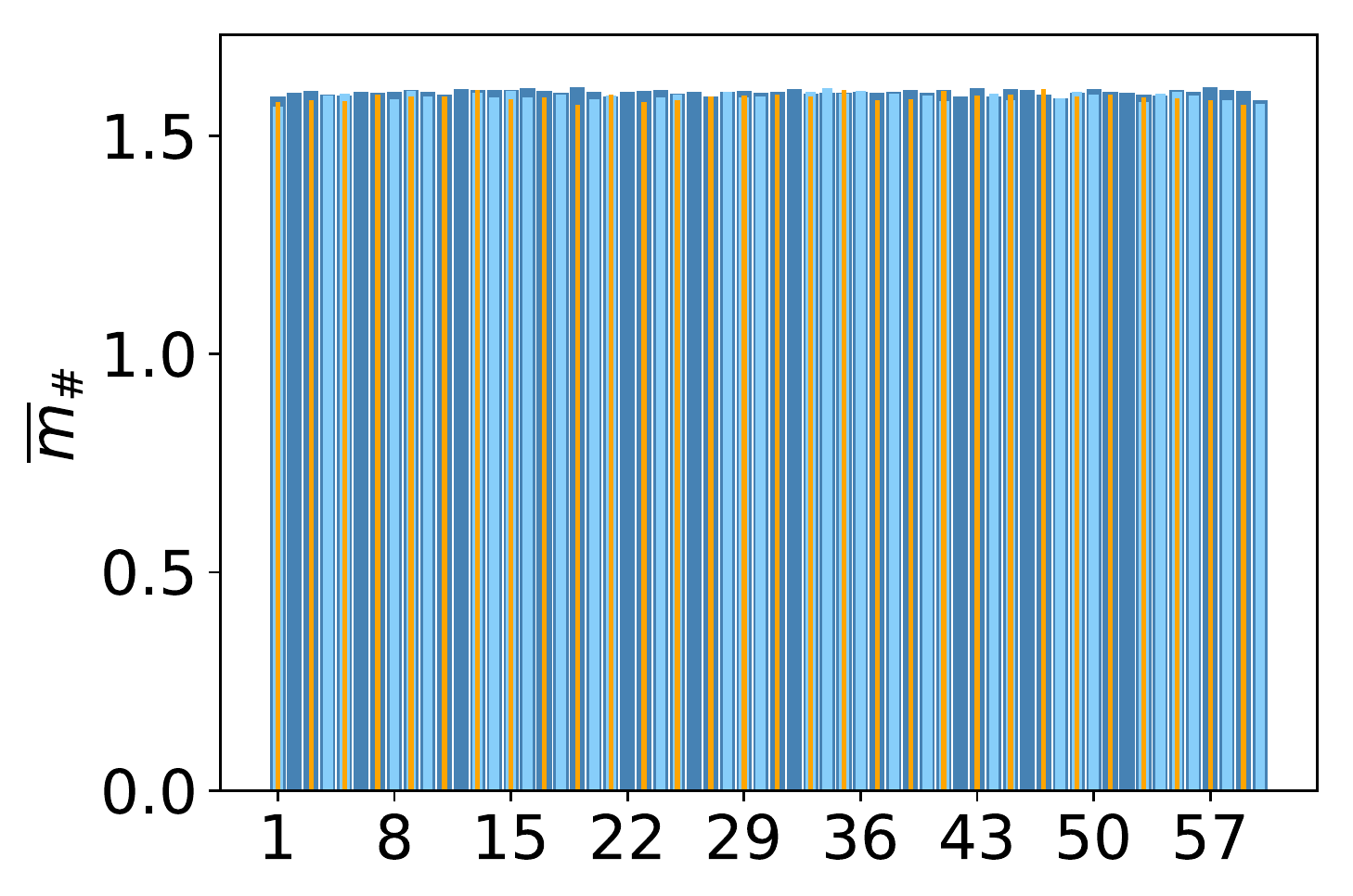}}
	\caption{Projection of the dipole moment of each magnetic bead in the direction of the applied field for MFs with \textit{plain} crosslinking. Along $x$-axis particle position is denoted. The sequencing is explained in the legend. In (a)--(d) $\mu^2 = 1$; in (e)--(h) $\mu^2=3$. In (a), (c), (e) and (g) $H=0.2$; in 
	(b), (d), (f) and (h) $H=6$.In (a), (b), (e) and (f) $L=20$; in (c), (d), (g) and (h) $L=60$.}
	\label{fig:magn_mom_pp-pcl}
\end{figure*}

The alignment with the field of the dipole moment of each magnetic particle might depend on its position along the MF backbone, due to differences in the local dipolar fields. We analyse this effect by calculating the projection of each dipole moment in the direction of the field. Fig.~ \ref{fig:magn_mom_pp-pcl} shows the results of this calculation as a function of the particle position within the filament for the case of \textit{plain} crosslinking, with $\mu^2=1$ in the upper row, $\mu^2=3$ in the lower one and two selected values of the field: $H=0.2$ (Figs.~\ref{fig:hist-l1-pcl-l20-h02}, \ref{fig:hist-l1-pcl-l60-h02} and \ref{fig:hist-l3-pcl-l20-h02}, \ref{fig:hist-l3-pcl-l60-h02}) and $H=6$ (Figs.~\ref{fig:hist-l1-pcl-l20-h60}, \ref{fig:hist-l1-pcl-l60-h60}, \ref{fig:hist-l3-pcl-l20-h60} and \ref{fig:hist-l3-pcl-l60-h60}). In these Figures one can notice that for $\mu^2=1$ (upper row), independently from the filament length, the influence of particle sequence on the magnetisation along the MF is very weak. It is, however, different for the lower row, where the increase in $\mu^2$ leads to local correlations between magnetic particles along the chain, which is particularly apparent at low $H$. Thus, in Fig.~\ref{fig:hist-l3-pcl-l60-h02} one can see correlated regions for each sequence and appreciate the difference in projection heights. As one can expect, for high fields this effect effectively vanishes.
\begin{figure*}
	\centering
\subfigure[]{\label{fig:hist-l1-ccl-l20-h02}\includegraphics[width=0.245 \textwidth]{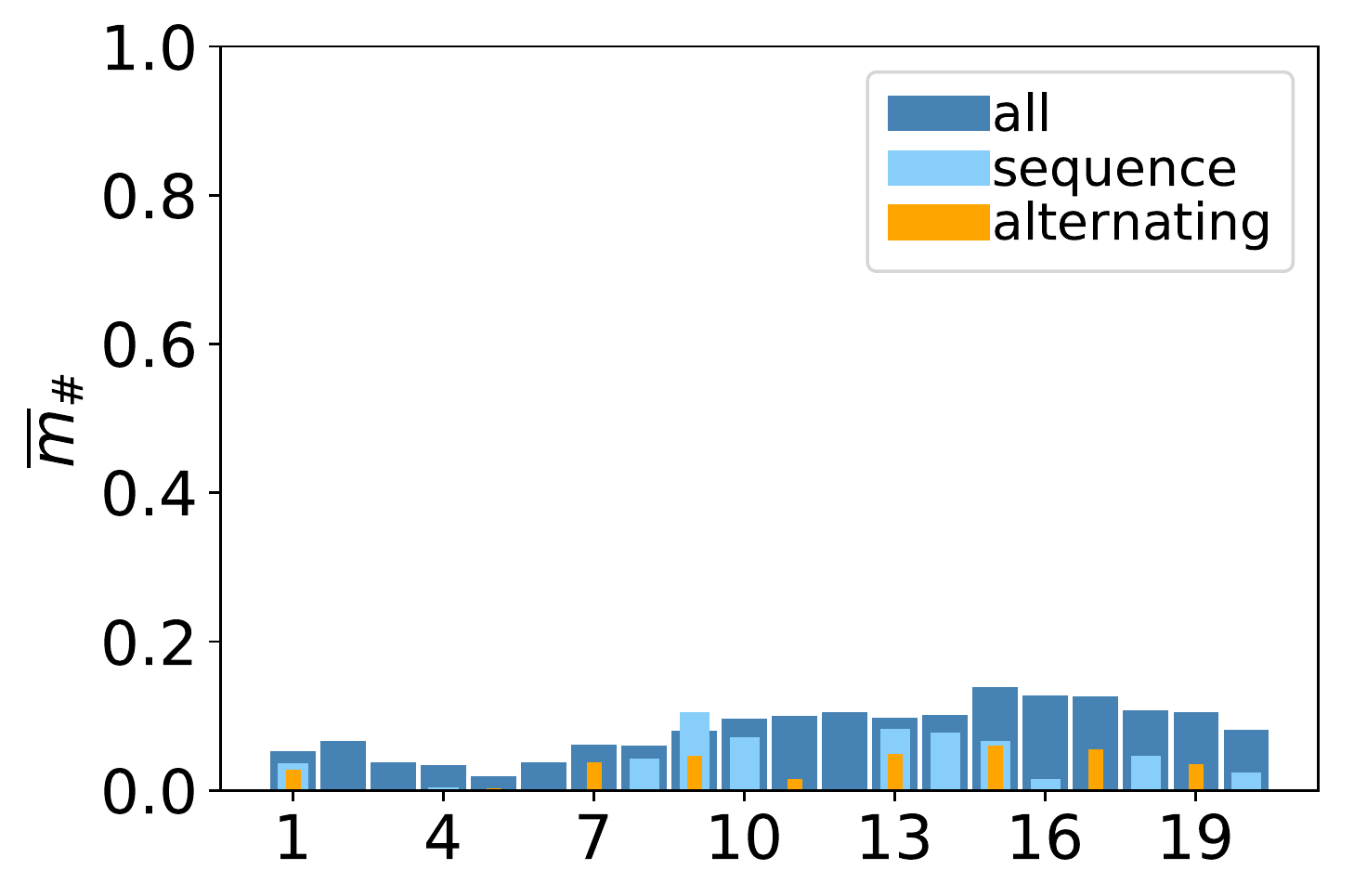}}
\subfigure[]{\label{fig:hist-l1-ccl-l20-h60}\includegraphics[width=0.245 \textwidth]{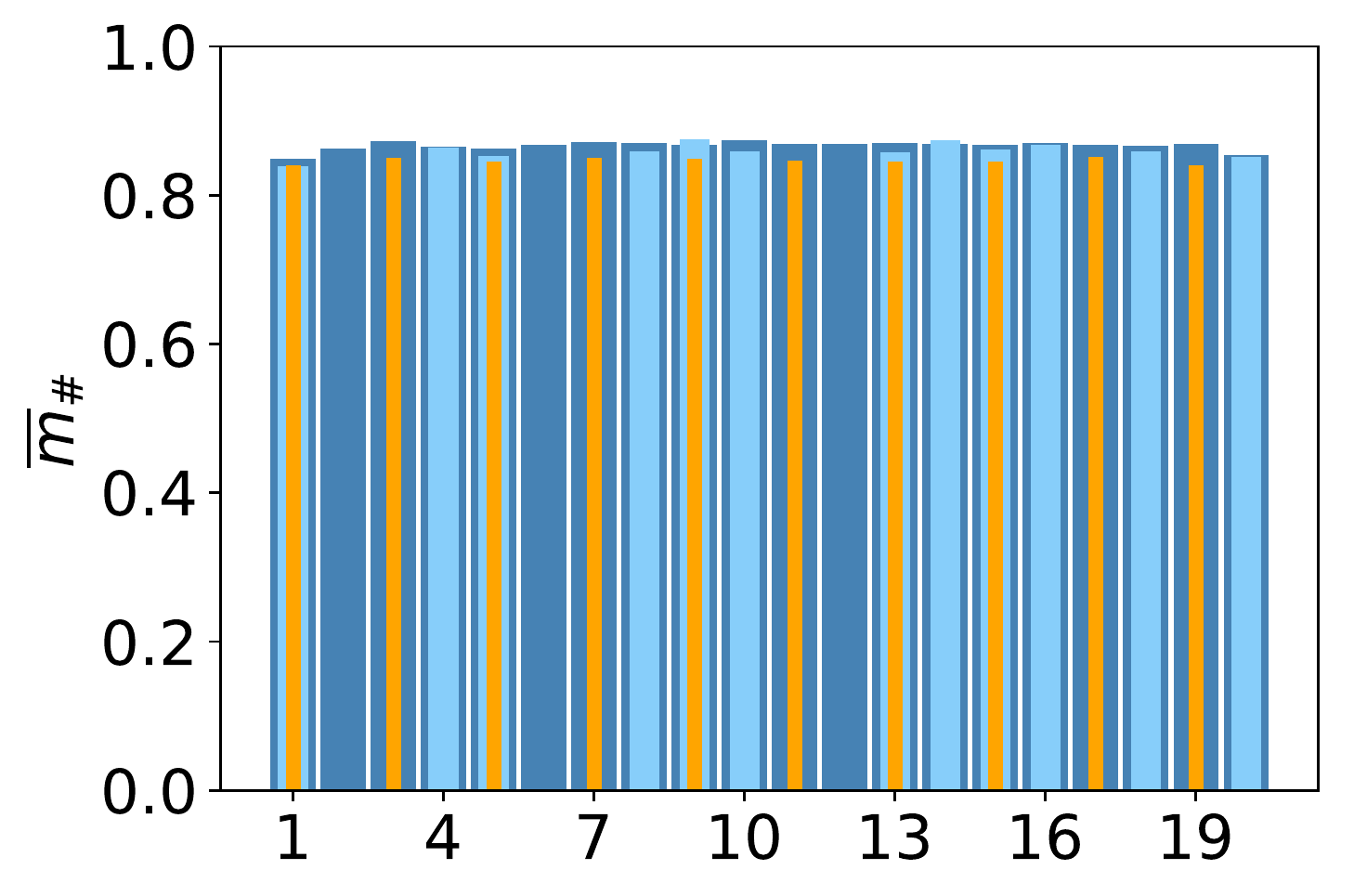}}
\subfigure[]{\label{fig:hist-l1-ccl-l60-h02}\includegraphics[width=0.245 \textwidth]{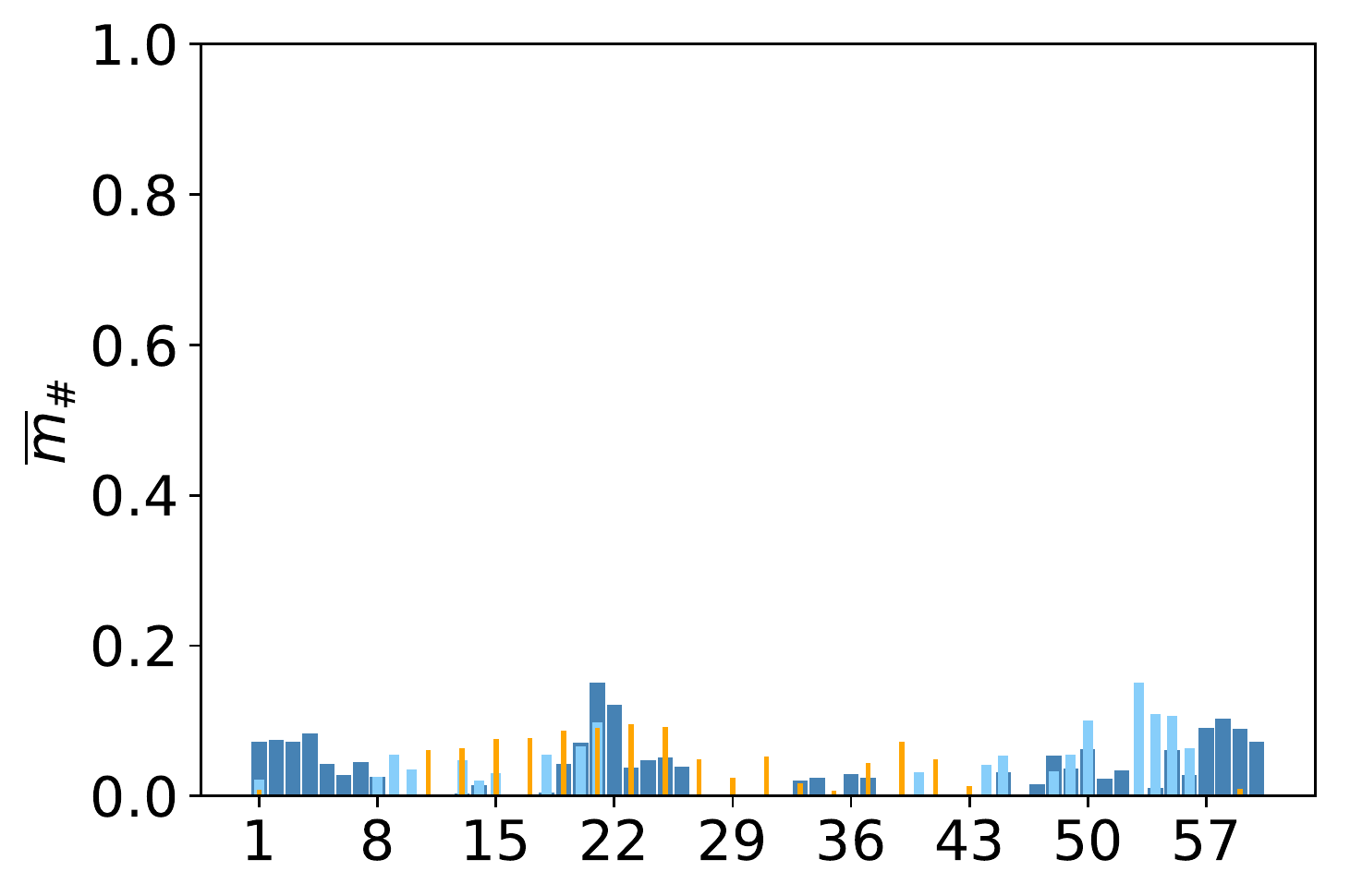}}
\subfigure[]{\label{fig:hist-l1-ccl-l60-h60}\includegraphics[width=0.245 \textwidth]{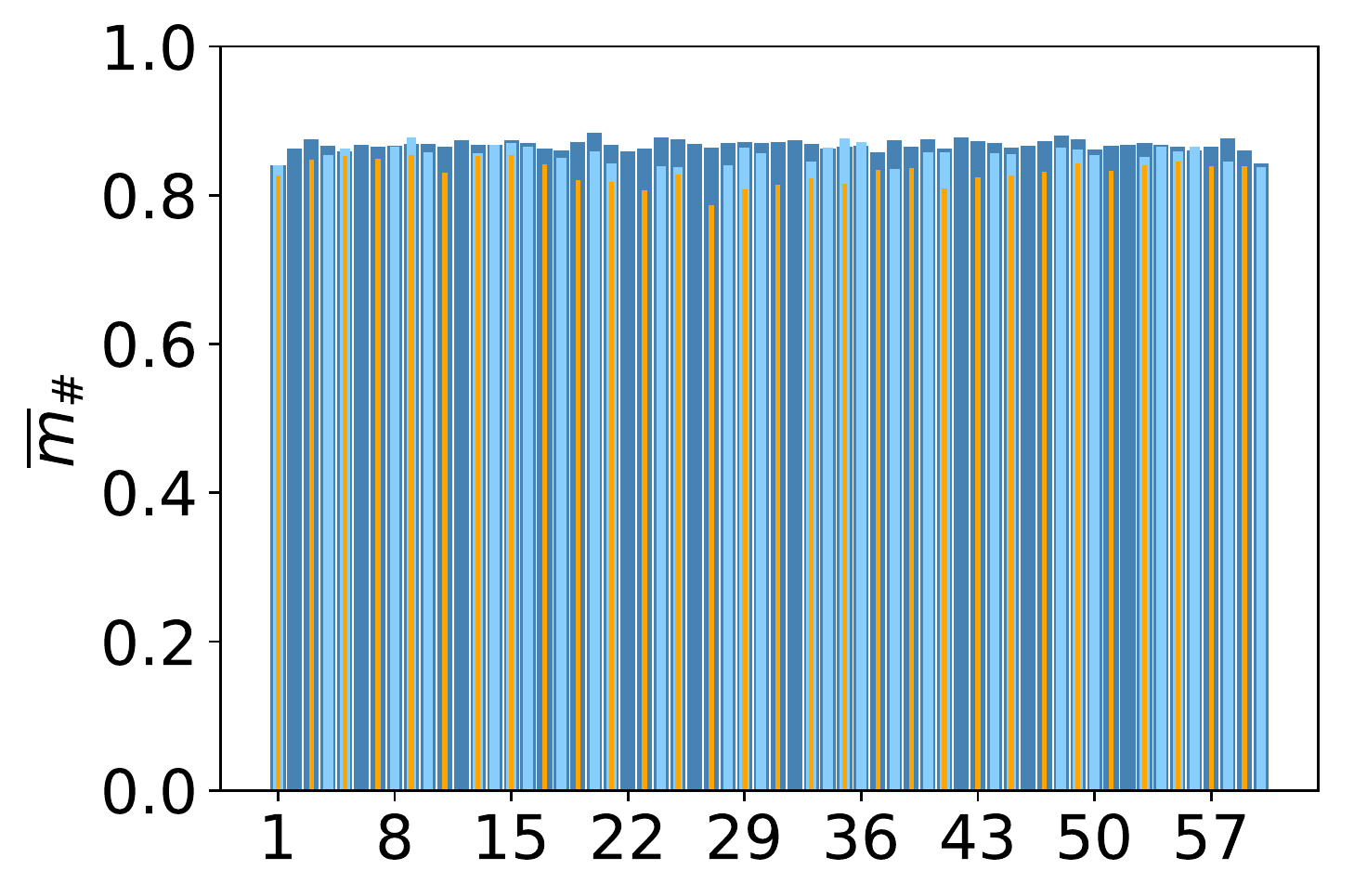}}
\subfigure[]{\label{fig:hist-l3-ccl-l20-h02}\includegraphics[width=0.245 \textwidth]{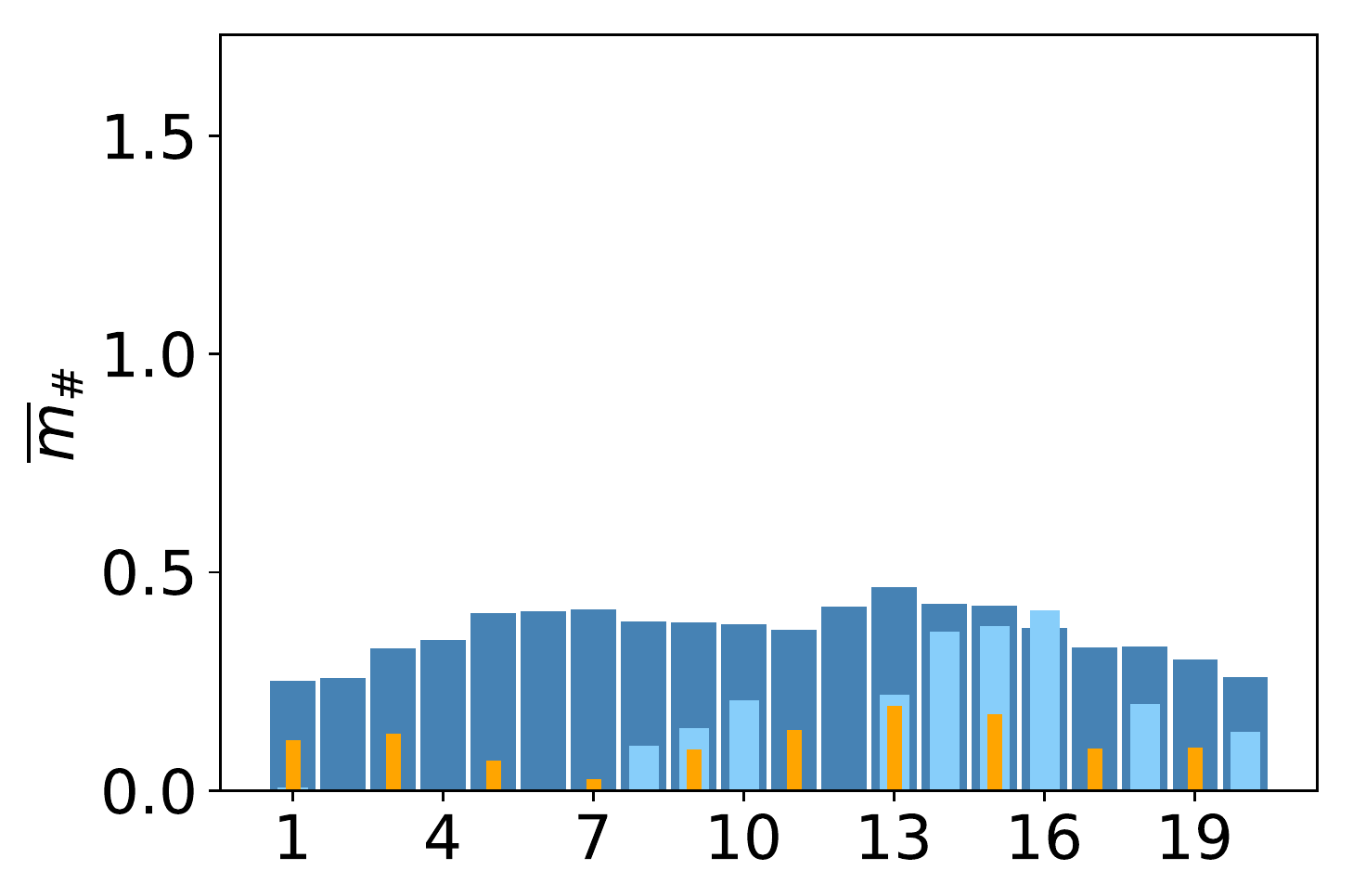}}
\subfigure[]{\label{fig:hist-l3-ccl-l20-h60}\includegraphics[width=0.245 \textwidth]{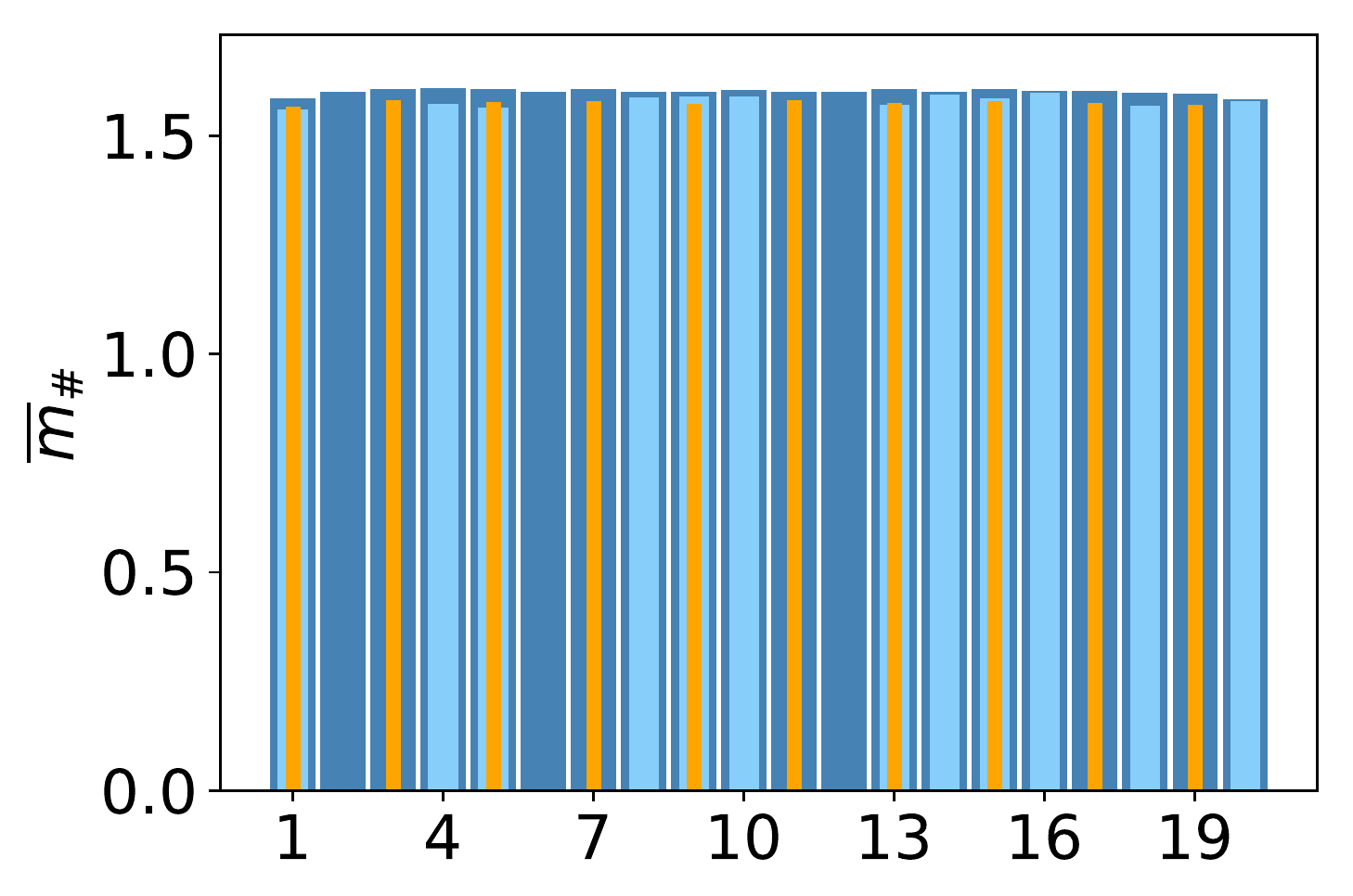}}
\subfigure[]{\label{fig:hist-l3-ccl-l60-h02}\includegraphics[width=0.245 \textwidth]{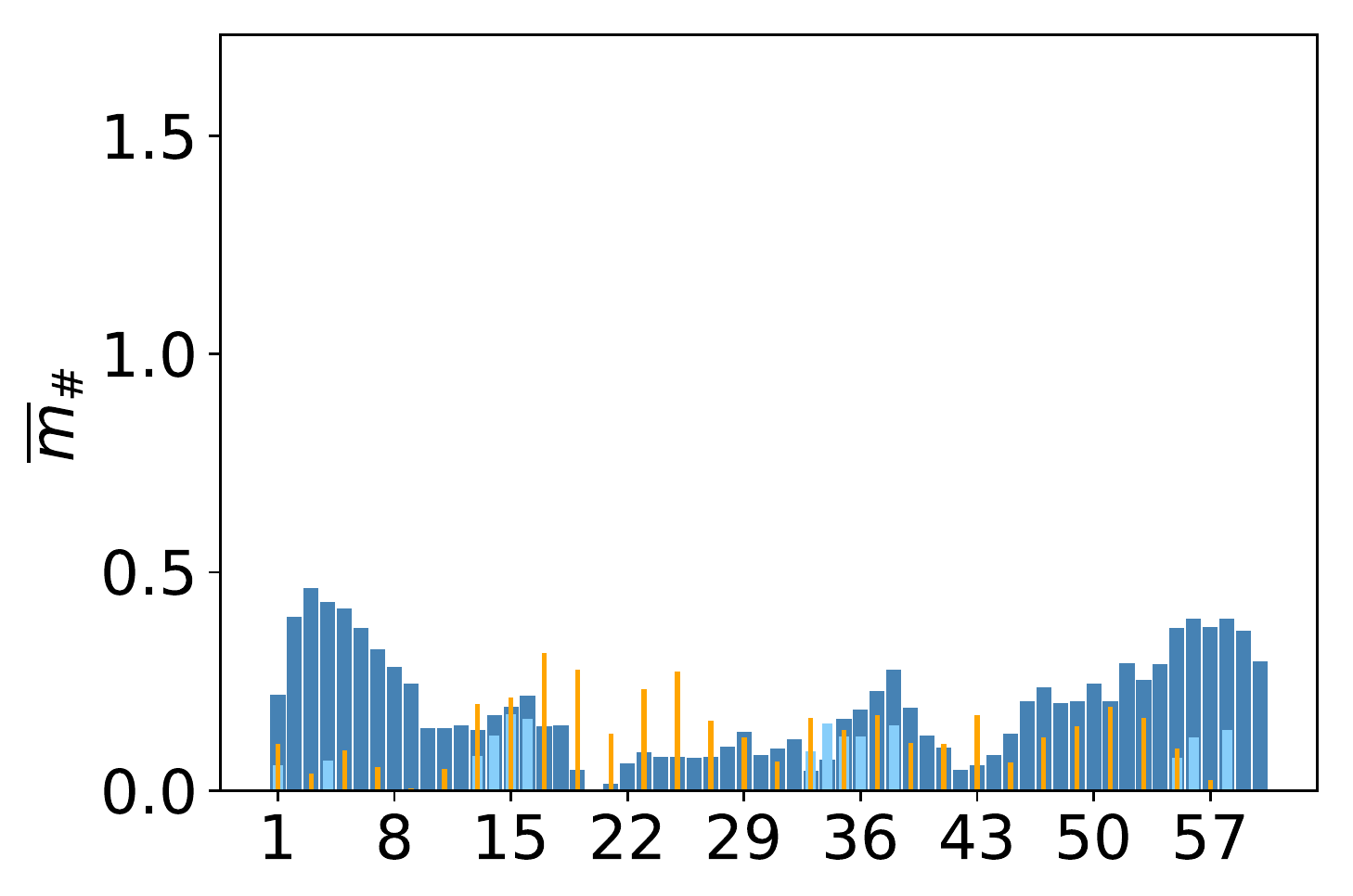}}
\subfigure[]{\label{fig:hist-l3-ccl-l60-h60}\includegraphics[width=0.245 \textwidth]{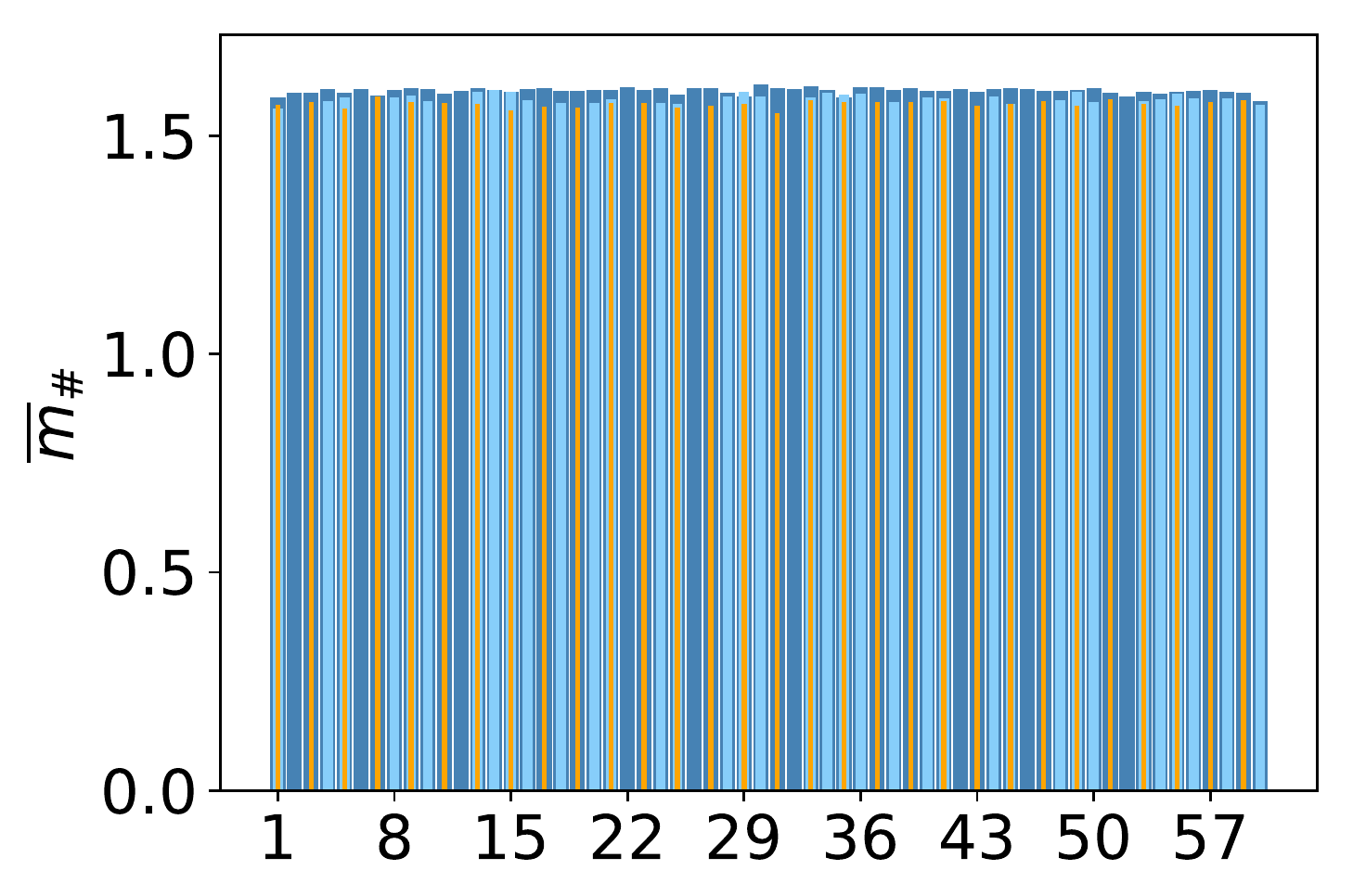}}
	\caption{Projection of the dipole moment of each magnetic bead in the direction of the applied field for MFs with \textit{constrained} crosslinking. Along $x$-axis, particle position is denoted. The sequencing is explained in the legend. In (a)--(d) $\mu^2 = 1$; in (e)--(h) $\mu^2=3$. In (a), (c), (e) and (g) $H=0.2$; in
	(b), (d), (f) and (h) $H=6$.In (a), (b), (e) and (f) $L=20$; in (c), (d), (g) and (h) $L=60$.}
	\label{fig:magn_mom_pp-ccl}
\end{figure*}

The picture changes when additional correlations are induced by crosslinking, as one can see in Fig.~\ref{fig:magn_mom_pp-ccl}. This figure has the same structure than Fig. \ref{fig:magn_mom_pp-pcl} but corresponds to \textit{constrained} crosslinking.
In contrast to the case above, the dipole projection in the direction of the field of differently positioned particles along the filament clearly depends on the particle sequence, even for high fields. It is worth noting that a larger dipole moment leads to the decrease of the aforementioned effect in high field, but appears to amplify it for low field.

\section{Conclusions}
In the present study we performed a thorough computer simulation analysis of the interplay between the crosslinking of magnetic filaments and their magnetic content, calculating their equilibrium magnetisation curves and positional distributions for the degree of alignment of their magnetic dipoles with the external field. For this, we considered two crosslinking mechanisms; one in which particle rotation is not penalised and only the interparticle distance is restricted by elastic springs (\textit{plain} crosslinking), and another in which not only the interparticle distance, but also any deviation of the magnetic moment from the backbone of the filament leads to an increase in the elastic energy of the connecting springs (\textit{constrained} crosslinking). In addition, we introduced three different sequences of magnetic/nonmagnetic particles along the filaments, differing in their relative number of magnetic particles and their positioning. We found that \textit{constrained} crosslinking, by inducing additional interparticle correlations, leads to larger differences in the magnetisation of filaments differing in their sequence type, as well as in the individual alignment of the dipoles with the field dependending on the particle position within the filament. 

\section{Acknowledgements}
This research has been supported by the Russian Science Foundation Grant No.19-12-00209.  Authors acknowledge support from the Austrian Research Fund (FWF), START-Projekt Y 627-N27.


\begin{thebibliography}{10}
\expandafter\ifx\csname url\endcsname\relax
  \def\url#1{\texttt{#1}}\fi
\expandafter\ifx\csname urlprefix\endcsname\relax\def\urlprefix{URL }\fi
\expandafter\ifx\csname href\endcsname\relax
  \def\href#1#2{#2} \def\path#1{#1}\fi

\bibitem{resler1964magnetocaloric}
E.~Resler, R.~Rosensweig, Magnetocaloric power, AIAA Journal 2~(8) (1964)
  1418--1422.

\bibitem{2009-odenbach}
S.~Odenbach (Ed.), Colloidal Magnetic Fluids, Vol. 763 of Lecture Notes in
  Physics, Springer-Verlag, Berlin Heidelberg, 2009.
\newblock \href {http://dx.doi.org/10.1007/978-3-540-85387-9}
  {\path{doi:10.1007/978-3-540-85387-9}}.

\bibitem{zrinyi1998kinetics}
M.~Zr{\i}nyi, D.~Szab{\'o}, H.-G. Kilian, Kinetics of the shape change of
  magnetic field sensitive polymer gels, Polymer Gels and Networks 6~(6) (1998)
  441--454.

\bibitem{Dreyfus_2005}
R.~Dreyfus, J.~Baudry, M.~L. Roper, M.~Fermigier, H.~A. Stone, J.~Bibette,
  \href{http://dx.doi.org/10.1038/nature04090}{Microscopic artificial
  swimmers}, Nature 437~(7060) (2005) 862--865.
\newblock \href {http://dx.doi.org/10.1038/nature04090}
  {\path{doi:10.1038/nature04090}}.
\newline\urlprefix\url{http://dx.doi.org/10.1038/nature04090}

\bibitem{2008-benkoski}
J.~J. Benkoski, S.~E. Bowles, R.~L. Jones, J.~F. Douglas, J.~Pyun, A.~Karim,
  Self-assembly of polymer-coated ferromagnetic nanoparticles into mesoscopic
  polymer chains, J Polym Sci, Part B: Polym Phy 46~(20) (2008) 2267--2277.
\newblock \href {http://dx.doi.org/10.1002/polb.21558}
  {\path{doi:10.1002/polb.21558}}.

\bibitem{1998-furst}
E.~M. Furst, C.~Suzuki, M.~Fermigier, A.~P. Gast, Permanently linked
  monodisperse paramagnetic chains, Langmuir 14~(26) (1998) 7334--7336.
\newblock \href {http://dx.doi.org/10.1021/la980703i}
  {\path{doi:10.1021/la980703i}}.

\bibitem{1999-furst}
E.~M. Furst, A.~P. Gast, Micromechanics of dipolar chains using optical
  tweezers, Phys Rev Lett 82~(20) (1999) 4130--4133.
\newblock \href {http://dx.doi.org/10.1103/PhysRevLett.82.4130}
  {\path{doi:10.1103/PhysRevLett.82.4130}}.

\bibitem{WANG_2011}
H.~WANG, Y.~YU, Y.~SUN, Q.~CHEN,
  \href{http://dx.doi.org/10.1142/S1793292011002305}{Magnetic nanochains: A
  review}, Nano 06~(01) (2011) 1--17.
\newblock \href {http://dx.doi.org/10.1142/s1793292011002305}
  {\path{doi:10.1142/s1793292011002305}}.
\newline\urlprefix\url{http://dx.doi.org/10.1142/S1793292011002305}

\bibitem{wang2014multifunctional}
H.~Wang, A.~Mararenko, G.~Cao, Z.~Gai, K.~Hong, P.~Banerjee, S.~Zhou,
  Multifunctional 1d magnetic and fluorescent nanoparticle chains for enhanced
  mri, fluorescent cell imaging, and combined photothermal/chemotherapy, ACS
  applied materials \& interfaces 6~(17) (2014) 15309--15317.

\bibitem{cebers2016flexible}
A.~Cebers, K.~Erglis, Flexible magnetic filaments and their applications,
  Advanced Functional Materials 26~(22) (2016) 3783--3795.

\bibitem{cai2018fluidic}
G.~Cai, S.~Wang, L.~Zheng, J.~Lin, A fluidic device for immunomagnetic
  separation of foodborne bacteria using self-assembled magnetic nanoparticle
  chains, Micromachines 9~(12) (2018) 624.

\bibitem{2005-dreyfus}
R.~Dreyfus, J.~Baudry, M.~L. Roper, M.~Fermigier, H.~A. Stone, J.~Bibette,
  Microscopic artificial swimmers, Nature 437~(7060) (2005) 862--865.
\newblock \href {http://dx.doi.org/10.1038/nature04090}
  {\path{doi:10.1038/nature04090}}.

\bibitem{2008-erglis-mh}
K.~\={E}rglis, L.~Alberte, A.~C\={e}bers, Thermal fluctuations of non-motile
  magnetotactic bacteria in ac magnetic fields, Magnetohydrodynamics 44~(3)
  (2008) 223--236.

\bibitem{fayol2013use}
D.~Fayol, G.~Frasca, C.~Le~Visage, F.~Gazeau, N.~Luciani, C.~Wilhelm, Use of
  magnetic forces to promote stem cell aggregation during differentiation, and
  cartilage tissue modeling, Advanced Materials 25~(18) (2013) 2611--2616.

\bibitem{gerbal2015refined}
F.~Gerbal, Y.~Wang, F.~Lyonnet, J.-C. Bacri, T.~Hocquet, M.~Devaud, A refined
  theory of magnetoelastic buckling matches experiments with ferromagnetic and
  superparamagnetic rods, Proceedings of the National Academy of Sciences
  112~(23) (2015) 7135--7140.

\bibitem{evans2007magnetically}
B.~Evans, A.~Shields, R.~L. Carroll, S.~Washburn, M.~Falvo, R.~Superfine,
  Magnetically actuated nanorod arrays as biomimetic cilia, Nano letters 7~(5)
  (2007) 1428--1434.

\bibitem{erglis2011three}
K.~{\=E}rglis, R.~Livanovi{\v{c}}s, A.~C{\=e}bers, Three dimensional dynamics
  of ferromagnetic swimmer, Journal of Magnetism and Magnetic Materials
  323~(10) (2011) 1278--1282.

\bibitem{2003-cebers}
A.~C\=ebers, Dynamics of a chain of magnetic particles connected with elastic
  linkers, J. Phys.: Condens. Matter 15~(15) (2003) S1335.
\newblock \href {http://dx.doi.org/10.1088/0953-8984/15/15/303}
  {\path{doi:10.1088/0953-8984/15/15/303}}.

\bibitem{kuznetsov2019equilibrium}
A.~A. Kuznetsov, Equilibrium properties of magnetic filament suspensions,
  Journal of Magnetism and Magnetic Materials 470 (2019) 28--32.

\bibitem{2013-arnold}
A.~Arnold, O.~Lenz, S.~Kesselheim, R.~Weeber, F.~Fahrenberger, D.~Roehm,
  P.~Ko{\v{s}}ovan, C.~Holm, Espresso 3.1: Molecular dynamics software for
  coarse-grained models, in: M.~Griebel, M.~A. Schweitzer (Eds.), Meshfree
  Methods for Partial Differential Equations VI, Springer Berlin Heidelberg,
  Berlin, Heidelberg, 2013, pp. 1--23.

\bibitem{weeks1971role}
J.~D. Weeks, D.~Chandler, H.~C. Andersen, Role of repulsive forces in
  determining the equilibrium structure of simple liquids, The Journal of
  chemical physics 54~(12) (1971) 5237--5247.

\bibitem{2015-sanchez-sm1}
P.~A. S\'anchez, J.~J. Cerd\`a, T.~Sintes, A.~O. Ivanov, S.~S. Kantorovich, The
  effect of links on the interparticle dipolar correlations in supramolecular
  magnetic filaments, Soft Matter 11 (2015) 2963--2972.
\newblock \href {http://dx.doi.org/10.1039/C5SM00172B}
  {\path{doi:10.1039/C5SM00172B}}.

\bibitem{ivanov2001magnetic}
A.~O. Ivanov, O.~B. Kuznetsova, Magnetic properties of dense ferrofluids: An
  influence of interparticle correlations, Physical Review E 64~(4) (2001)
  041405.

\end{thebibliography}
\end{document}